\documentclass[a4paper,headings=standardclasses, headings=big]{scrartcl}
\usepackage{url}
\usepackage[english]{babel}

\usepackage{amsmath}
\usepackage{amssymb}
\usepackage{amsthm}
\usepackage{graphicx}
\usepackage{mathtools}

\graphicspath{{fig/}}
\theoremstyle{plain} 

\newtheorem*{rem*}{Remark}

\DeclareMathOperator{\diver}{div}
\DeclareMathOperator{\grad}{grad}
\DeclareMathOperator{\Diver}{Div}
\DeclareMathOperator{\Grad}{Grad}

\DeclareMathOperator{\tr}{tr}

\DeclareMathOperator{\curl}{curl}
\DeclareMathOperator{\Curl}{Curl}

\usepackage{mathtools}
\usepackage{pgfplots}
\pgfplotsset{/pgf/number format/use comma,compat=newest}

\renewcommand\epsilon{\varepsilon}
\newcommand{\R}{\mathbb{R}}

\renewcommand{\d}{\mathrm{d}}

\newcommand{\sistema}[1]{\left\{\begin{aligned}#1\end{aligned}\right.}
\newcommand{\vect}[1]{\boldsymbol{#1}}
\newcommand{\tens}[1]{\mathsf{#1}}

\title{Tunable morphing of electroactive dielectric-elastomer balloons}
\author{Yipin Su$^{1,\,*}$, Davide Riccobelli$^{1,\,*}$, Yingjie Chen$^{2,\,*}$,\\ Weiqiu Chen$^{2,\,3}$, Pasquale Ciarletta$^{1}$\\
$^{1}$ MOX -- Dipartimento di Matematica, Politecnico di Milano,\\Piazza Leonardo da Vinci 32, Milan 20133, Italy\\
$^{2}$ Department of Engineering Mechanics, Zhejiang University,\\Hangzhou 310027, PR China\\
$^{3}$Shenzhen Research Institute of Zhejiang University,\\Shenzhen 518057, PR China\\
$^*$ These authors equally contributed to the work.
}

\graphicspath{{fig/}}

\begin{document}
\maketitle

\begin{abstract}
Designing smart devices with tunable shapes has important applications in industrial manufacture. In this paper, we investigate the nonlinear deformation and the morphological transitions between buckling, necking, and snap-through instabilities of layered DE balloons in response to an applied radial voltage and an inner pressure.  We propose a general mathematical theory of nonlinear electro-elasticity able to account for finite inhomogeneous strains provoked by the electro-mechanical coupling.
We investigate the onsets of morphological transitions of the spherically symmetric balloons using the surface impedance matrix method. Moreover, we study the nonlinear evolution of the bifurcated branches through finite element numerical simulations.  Our analysis demonstrates the possibility to design tunable DE spheres, where the onset of buckling and necking can be controlled by geometrical and mechanical properties of the passive elastic layers. Relevant applications include soft robotics and mechanical actuators. 
\end{abstract}

\section{Introduction}
Dielectric elastomers (DEs) are soft  smart materials capable of performing large  deformations in fast response to electrical stimuli. In the last decades, they have 
 attracted considerable attention, from both academia and industry, for many  applications at the core of modern technologies, such as  soft robots, artificial muscles,
 actuators and energy harvesters \cite{Kim07, Halloran2008}. A typical DE actuator consists of a soft elastomer sandwiched between two compliant electrodes. The actuator deforms when subject to a voltage along the thickness direction, accompanied by a reduction in the thickness and an expansion in the area \cite{Suo08, Su2019a}. The DE balloon is widely used as a device configuration for its suitability to enhance the electric-induced deformation and for its versatility in industrial applications \cite{Huang2013, Bortot17}. In some practical applications, the DE devices should be insulated from their surroundings. For example,   a wearable device is capable of delivering haptic information by adding a soft elastic insulation layer  outside the DE actuator,   protecting human skin, enhancing breathability and preventing slippage \cite{Lee2022}.

\begin{figure}[t!]
\centering
\includegraphics[width=1\textwidth]{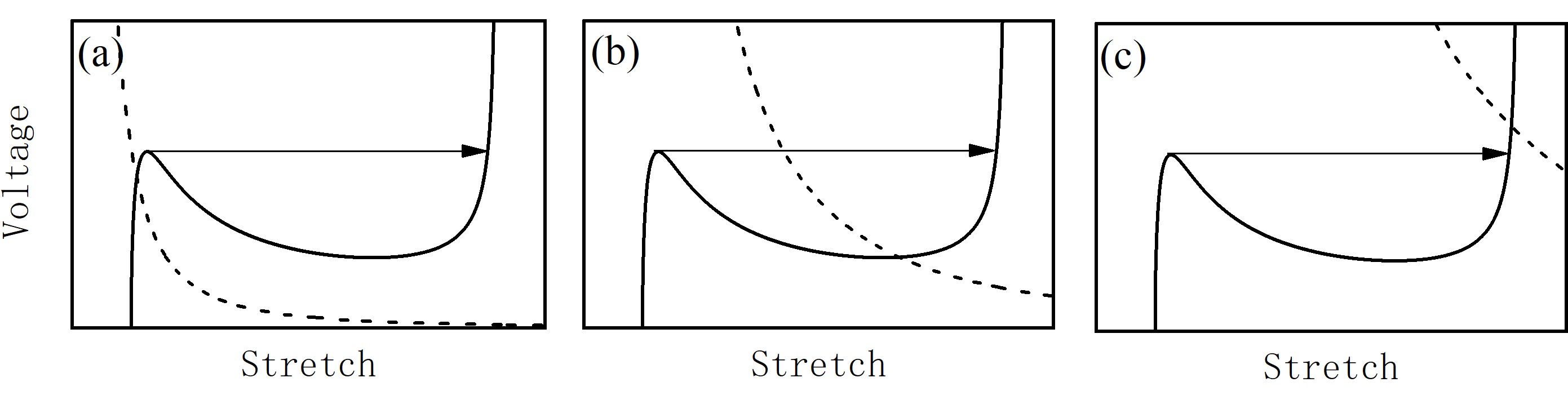}
\caption{
Nonlinear responses of planar DE actuators subject to a voltage through the thickness: (a) Electrical breakdown occurs prior to the onset of snap-through instability (type I); (b) Snap-through
instability induces electrical breakdown of the material (type II); (c) The DEs survives the snap-through instability (type III). The solid and dashed curves correspond to the voltage-stretch loading curve and the electric breakdown curve of the materials, respectively. The arrow represents the snap-through path of the material during the deformation.
}
\label{figure1}
\end{figure}

 A clear advantage of using  DEs is the N-shaped constitutive curve between the applied  voltage and the resulting stretch, which characterizes the  so-called \textit{snap-through instability} \cite{Zhu10, Su2018}. It consists of a sudden increase of the resulting  strain as the applied voltage reaches the local maximum value,  as presented in Fig. \ref{figure1}. Taking advantage of this feature, a giant areal strain can be obtained, as large as 1600\% \cite{Li2013, Godaba2014}.  

Due to their geometric and constitutive nonlinearities,  as well as to the multi-physics coupling, DE devices can be modeled as boundary value problems (BVPs) that are generally  difficult to solve. Multiple morphological transitions can occur due to the presence of many stationary points of the electro-mechanical functional of interest. For instance, necking (i.e. the local thinning of an elastic body under tension) was initially identified as a precursor of structural failure of DE devices, which should be avoided \cite{Chen2017}. On the contrary, more recent studies have reported that compressive buckling can be positively  used to control shape and surface patterns in functional devices and biological tissues \cite{Pang2020, Xu2020}.

In this respect, the buckling of nonlinear elastic, thick-walled balloons has been extensively studied in the past \cite{Haughton1978, haughton1980post, debotton2013axisymmetric}
So far, some efforts have been devoted in the literature to the study of  mechanical behaviors of  DE balloons.  For example, Alibakhshi et al. studied the nonlinear vibration and stability of a dielectric elastomer balloon based on a strain-stiffening model \cite{Alibakhshi2023}. Jin and Huang investigated the random response of dielectric elastomer balloons disturbed by electrical or mechanical fluctuation \cite{Jin2017}. Sharma et al. developed an energy-based method for estimating the dynamic pull-in instability parameters of the DE balloon actuator undergoing homogeneous deformation in \cite{Sharma2018}. Liang and Cai proposed a study of shape bifurcation of a spherical dielectric elastomer balloon subject to internal pressure and electric voltage \cite{Liang2015}. Xie et al. studied the bifurcation of a dielectric elastomer balloon under pressurized inflation and electric actuation \cite{Xie2016}. Rudykh et al. \cite{Rudykh2009} investigated the response of electroactive balloons subject to coupled electromechanical stimuli. The inflation deformation of an electroelastic spherical shell subject to an internal pressure and a radial voltage was examined in \cite{Dorfmann2014}. Mao et al. conducted a 3D analytical study of the small-amplitude free vibration of a SEA spherical balloon with radially inhomogeneous biasing fields \cite{Mao2019}. The bifurcation of finitely deformed thick-walled electroelastic spherical shells subject to a radial electric field was analyzed in \cite{Melnikov2020}.  
It should be noted that most existing works focused on monolayer DE balloons.  Recently, layered dielectric composites have gained more and more attention. Osman et al. proposed the approaches for preparing bilayered polydimethylsiloxane (PDMS) composite for dielectric elastomer applications. Kumar et al. theoretically studied the dynamic electromechanical behavior of multi-layered DE composites \cite{Kumar2023}. Su et al. proposed a dielectric-elastomer bilayer capable of smart bending deformation \cite{Su2020}. Four different criteria of multilayered soft dielectrics under plane-strain conditions were compared in \cite{Bertoldi2011}.  To the best knowledge of the authors, the only related work on the nonlinear response of  multi-layered DE balloons was proposed by \cite{Bortot17}. However, the influences of the layered configuration and the applied electro-mechanical stimuli on the morphological diagram  of layered DE balloons are still unknown.  By analyzing the electro-elastic behavior of a spherical piezoceramic sensor coated by a homogeneous protective layer, it was shown that the existence of the protective layer can prolong the effective working life of the piezoceramic sensor \cite{atashipour2016electro}. It is unclear whether or not this enhancement can happen in dielectric devices. 

Here, we propose a theoretical and numerical study  of an incompressible layered dielectric-elastic balloon subject to the combined action of electrical and mechanical loads. For simplification, we only consider type III DEs as defined in Fig. \ref{figure1}$c$, excluding the possibility of electric breakdown of the material before the onsets of the snap-through and the necking instabilities. We investigate the possibility to enhance and to control the electric-induced deformation field by coating an inactive elastic layer outside the DE balloon.

The article is structured as follows. In Section \ref{section-nonlinear}, we derive the governing equations for the inflation of a layered dielectric-elastic  balloon subject to an internal pressure and a radial voltage. We assume that the elastic and DE layers are perfectly bonded. In Section \ref{section-incremental}, we formulate the linear stability analysis of the radially symmetric solution, using the surface impedance matrix method to implement a robust numerical scheme for solving the linearised BVP for a generic constitutive function.  In Section \ref{section-energy},  we illustrate the solution obtained by assuming a  Gent constitutive response. In Section \ref{section-numerical},  we compare the numerical results obtained for elastic monolayer, DE monolayer and bi-layered dielectric-elastic balloons. In Section \ref{section-post-bifurcation}, we present a post-buckling analysis of the devices  by means of a finite element approximation of the nonlinear problem.  In Section \ref{section-conclusion}, we finally  discuss the relevance of our results for enabling a novel design strategy for tuning shapes of DE devices, together with a few concluding remarks.

\section{The electro-mechanical model}
\label{section-nonlinear} 
In this section, we  define the nonlinear BVP describing the response of a spherical dielectric-elastic balloon to electro-mechanical stimuli, deriving its  radially symmetric solution.
\subsection{Nonlinear boundary value problem}

\begin{figure}[t!]
\centering
\includegraphics[width=0.8\textwidth]{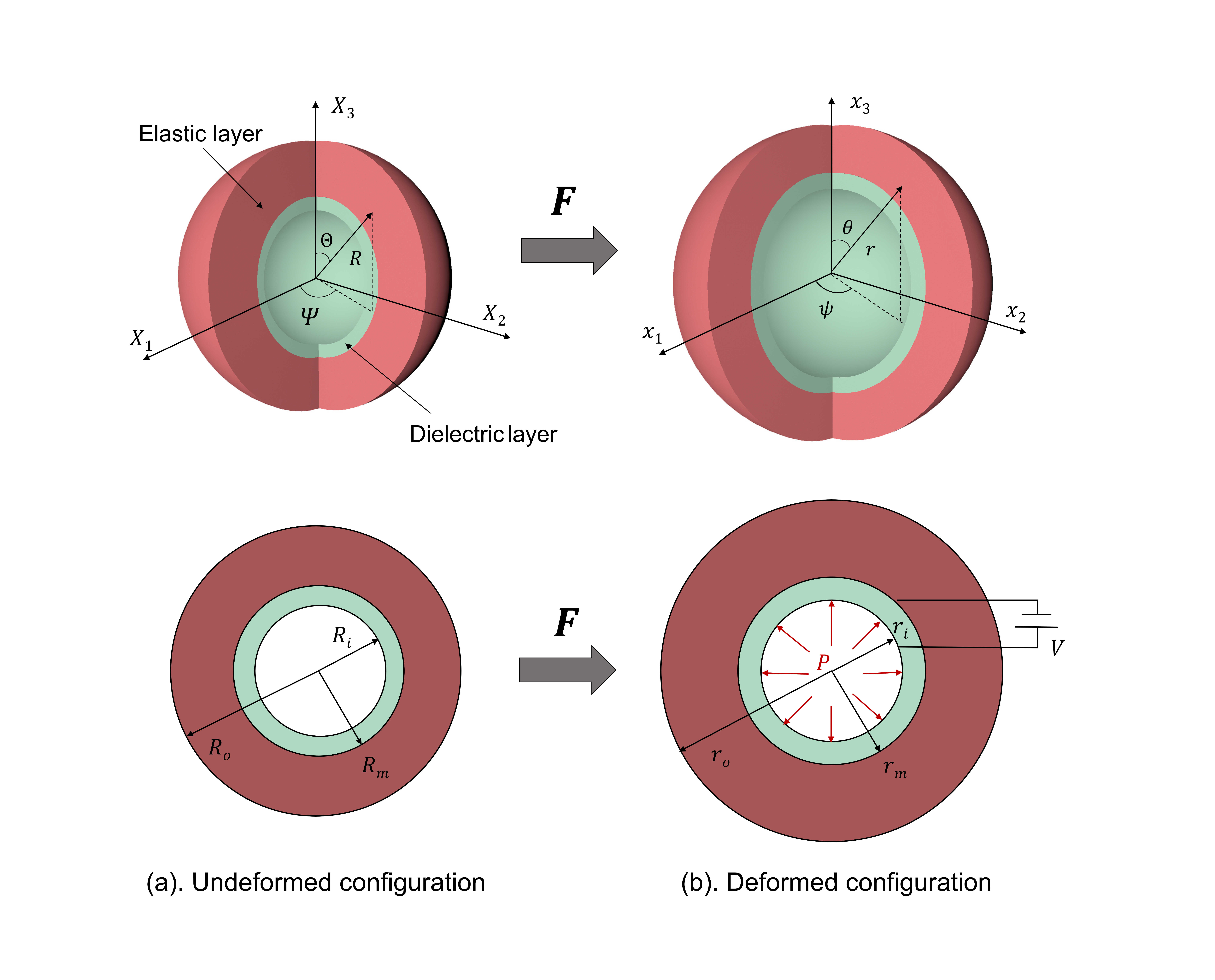}
\caption{
Sketch of a layered dielectric-elastic balloon subject to a radial voltage $V$ through the DE layer and an internal pressure $P$ at the inner surface: (a) undeformed and (b) deformed configurations. The upper row depicts the three-dimensional domains with the respective spherical coordinate systems,  and the lower row shows the corresponding in-plane cross sections with geometrical and electro-mechanical loading parameters.
}
\label{figure2}
\end{figure}

Let us consider a thick-walled bi-layered spherical balloon that occupies the domain $\mathcal{B}_0\subset\mathbb{R}^3$ in its reference configuration. Specifically $\mathcal{B}_0=\mathcal{B}^d_0\cup\mathcal{B}^e_0$, where
\begin{equation}
\begin{aligned}
\mathcal{B}^d_0 &= \{\vect{X}\in\mathbb{R}^3\;|\;R_i<\|\vect{X}\|\leq R_m\},\\
\mathcal{B}^e_0 &= \{\vect{X}\in\mathbb{R}^3\;|\;R_m<\|\vect{X}\|\leq R_o\},
\end{aligned}
\end{equation}
as illustrated in Fig. \ref{figure2}. We assume that the inner and outer layers are made of DE and elastic elastomers, respectively. Throughout the paper, we denote  the quantities related to the inner, interfacial and outer surfaces  by the subscripts ${\left(  \bullet  \right)_i}$, ${\left(  \bullet  \right)_m}$ and ${\left(  \bullet  \right)_o}$, respectively, and the quantities related to the DE and elastic layers  by the superscripts ${\left(  \bullet  \right)^d}$ and ${\left(  \bullet  \right)^e}$, respectively. $\vect{X}$ and $\vect{x} = \vect{\chi}(\vect{X})$ are the reference and the actual position vectors, respectively, with $\vect{\chi}:\mathcal{B}_0\rightarrow \mathbb{R}^3$ being the mapping from the reference configuration to the actual configuration $\mathcal{B} = \vect{\chi}(\mathcal{B}_0)$. Then the displacement field is $\vect{u}= \vect{x}-\vect{X}$.

Let $(\vect{e}_R,\,\vect{e}_\Psi,\,\vect{e}_\Theta)$ and $(\vect{e}_r,\,\vect{e}_\psi,\,\vect{e}_\theta)$ be the corresponding spherical orthonormal bases in the reference and actual configurations, respectively. Moreover, let $(R,\,\Psi,\,\Theta)$ and $(r,\,\psi,\,\theta)$ be the spherical coordinates in the reference and actual frames, respectively, so that
\begin{equation}
\sistema{
&\vect{X} = (X_1,\,X_2,\,X_3)=(R\sin\Theta\cos\Psi,\,R\sin\Theta\sin\Psi,\,R\cos\Theta),\\
&\vect{x}= (x_1,\,x_2,\,x_3)= (r\sin\theta\cos\psi,\,r\sin\theta\sin\psi,\,r\cos\theta).
}
\end{equation} 

We denote by $\tens{F} = \Grad\vect{\chi}$ the deformation gradient, where $\Grad$ is the material gradient operator. In the following, we use capital letters for all the differential operators referring to  the reference frame. We assume that both layers are incompressible, so that everywhere holds the constraint
\begin{equation}
\label{eq:inc}
\det\tens{F} = 1.
\end{equation}

We now introduce the true electric field $\vect{E}:\mathcal{B}\rightarrow\R^3$ and the true electric displacement field $\vect{D}:\mathcal{B}\rightarrow\R^3$. Their Lagrangian counterparts are given by
\begin{equation}
\vect{E}_L = \tens{F}^T \vect{E},\qquad\vect{D}_L = \tens{F}^{-1}\vect{D}.
\end{equation}
Accordingly, the Maxwell equations in the material frame read
\begin{equation}
\label{eq:maxwell}
\Curl \vect{E}_L = \vect{0},\qquad\Diver\vect{D}_L=0,
\end{equation}
or, equivalently, in the material setting
\begin{equation}
\curl \vect{E} = \vect{0},\qquad\diver\vect{D} = 0, 
\end{equation}
where $\diver$ and $\curl$ are the spatial divergence and curl operators, respectively,  and $\Diver$ and $\Curl$ their material counterparts.
The first equation is automatically satisfied if we introduce the electric potential $\phi$ such that:
\begin{equation}
\vect{E} = -\grad\phi.
\end{equation}

Both in $\mathcal{B}_0^d$ and  in $\mathcal{B}_0^e$, the nominal stress $\tens{S}$ must satisfy the balance equation
\begin{equation}
\label{eq:balance}
\Diver \tens{S} = \vect{0},
\end{equation}
while we require that the normal traction and displacement are continuous at the interface between the elastomer and the dielectric layers, that is
\begin{equation}
\label{eq:interface}
\lim_{R\rightarrow R^-_m} \tens{S}^T\vect{e}_R = \lim_{R\rightarrow R^+_m} \tens{S}^T\vect{e}_R.
\end{equation}

We postulate the existence of an energy density $W$ in the form
\begin{equation}
W = \left\{
\begin{aligned}
&W^d(\tens{F},\,\vect{D}_L) &&\text{in }\mathcal{B}_0^d,\\
&W^e(\tens{F}) &&\text{in }\mathcal{B}_0^e.
\end{aligned}
\right.
\end{equation}
The dielectric elastomer is assumed to be isotropic. In virtue of the representation theorem of isotropic functions,  the energy density can be generally written as a function of five invariants, namely $W^d = W^d(I_1,\,I_2,\,I_4,\,I_5,\,I_6)$, where
\begin{equation}
\label{eq:invariants}
\begin{gathered}
I_1 = \tr\tens{C},\qquad
I_2 = \frac{I_1^2-\tr (\tens{C}^2)}{2},\qquad
I_4 =\vect{D}_L\cdot\vect{D}_L,\\
I_5 =\vect{D}_L\cdot\tens{C}\vect{D}_L,\qquad
I_6 =\vect{D}_L\cdot\tens{C}^2\vect{D}_L,
\end{gathered}
\end{equation}
where $\tens{C}$ is the right Cauchy--Green tensor, given by $\tens{C} = \tens{F}^T\tens{F}$. The elastomer is also isotropic and therefore $W^e = W^e(I_1,\,I_2)$.

By standard thermo-mechanical considerations \cite{dorfmann2014nonlinear}, the nominal stress and the Lagrangian electric field can be expressed as
\begin{equation}
    \label{eq:PE_W}
\tens{S} = \frac{\partial W}{\partial \tens{F}} - p \tens{F}^{-1},\qquad \vect{E}_L = \frac{\partial W}{\partial \vect{D}_L},
\end{equation}
where $p$ is a Lagrange multiplier that enforces the incompressibility constraint Eq.~\eqref{eq:inc}. For later convenience, we also introduce the push-forward of the nominal stress in the actual configuration, i.e. the Cauchy stress tensor
\begin{equation}
\tens{T} = \tens{F}\tens{S}.
\end{equation}

The Maxwell equation \eqref{eq:maxwell} and the balance equation \eqref{eq:balance} are complemented by interface and boundary conditions. We assume that the inner and the outer surfaces of the DE layer are coated with compliant electrodes, so that the potential difference between the surfaces of the dielectric is a fixed applied voltage $V$, namely
\begin{equation}
\label{eq:boundary1}
\Delta \phi = \phi(R_m) - \phi(R_i)= V.
\end{equation}
Furthermore, we assume that the inner surface of the bilayer is subject to a pressure $P$, such that
\begin{equation}
\label{eq:boundary2}
\tens{S}^T\vect{N} = - P \tens{F}^{-T}\vect{N},
\end{equation}
where $\vect N$ is the outward normal vector.

In the following, we derive the radially symmetric solution of the BVP given by Eqs.~\eqref{eq:inc}-\eqref{eq:boundary2}.

\subsection{Radially symmetric solution}
We look for a particular solution of the BVP using the ansatz
\begin{equation}
    \label{eq:symmetry}
r=r(R),\qquad \theta = \Theta,\qquad \psi = \Psi,\qquad \phi=\phi(R).
\end{equation}
We introduce the spatial inner, interfacial and outer radii $r_i=r(R_i)$, $r_m=r(R_m)$ and $r_o=r(R_o)$, respectively. Using spherical coordinates, the deformation gradient reads
\begin{equation}
\tens{F}=\lambda_r\vect{e}_r\otimes\vect{E}_R+\lambda_\theta\vect{e}_\theta\otimes\vect{E}_\Theta+\lambda_\psi\vect{e}_\psi\otimes\vect{E}_\Psi=\frac{\d r}{\d R}\vect{e}_r\otimes\vect{E}_R+\frac{r}{R}\left(\vect{e}_\theta\otimes\vect{E}_\Theta+\vect{e}_\psi\otimes\vect{E}_\Psi\right),
\end{equation}
where $\lambda_j\ (j=r, \theta, \psi)$ is the stretch at $j-$direction of the balloon.

Using the incompressibility assumption Eq.~\eqref{eq:inc} in its local and global forms,  we get
\begin{equation}
    \label{deformation_equation}
\frac{\d r}{\d R}\frac{r^2}{R^2}=1 \quad\Longrightarrow\quad r(R) = \sqrt[3]{R^3-R_i^3+r_i^3}.
\end{equation}

We denote  the hoop stretch as $\lambda = r/R$, so that
\begin{equation} \label{stretch}
\lambda_r = \frac{1}{\lambda^2},\quad\lambda_\theta=\lambda_\psi=\lambda.
\end{equation}
From Eq.~\eqref{deformation_equation}, the following relationships hold between the stretches at the inner, interfacial and outer surfaces $\lambda_i=r_i/R_i$, $\lambda_m=r_m/R_m$ and $\lambda_o=r_o/R_o$ of the balloon,
\begin{equation} \label{stretch-relationship}
\begin{array}{l}
{\lambda _m} = {\left[ {1 + {(t^d)^3}\left( {\lambda _i^3 - 1} \right)} \right]^{1/3}},  \qquad {\lambda _o} = {\left[ {1 + {(t^e)^3}\left( {\lambda _m^3 - 1} \right)} \right]^{1/3}},
\end{array}
\end{equation}
where  $t^d= {R_i}/{R_m}$ and $t^e= {R_m}/{R_o}$ are the aspect ratios of the DE and elastic layers, respectively. 
We remark that the limit cases $t^d \to 1$ and $t^e \to 1$ correspond to monolayer elastic and DE balloons, respectively. 

The radial electric field  generated by the applied potential difference $V$  only exists in the DE layer. According to our ansatz, the nominal electric field and nominal electric displacement have the forms
\begin{equation}
 \vect{E}_L= E_R\vect{e}_R,
\quad
 \vect{D}_L=D_R\vect{e}_R,
\end{equation} 
where $E_R$ and $D_R$ are the only non-zero components of the nominal electric field and nominal electric displacement, respectively. 
The corresponding true electric field and true electric displacement are
\begin{equation}
 \vect{E}=\tens{F} ^{- T} \vect{E}_L= E_r \vect{e}_r
\qquad
 \vect{D}=\tens{F} \vect{D}_L= D_r \vect{e}_r,
\end{equation} 
where $E_r=\lambda^2E_R$ and $D_r=\lambda^{-2}D_R$ are the only non-zero components of the true electric field and true electric displacement, respectively.

From Eqs.~\eqref{eq:invariants} and~\eqref{stretch}, we obtain the following expressions of the invariants
\begin{equation}\label{invariants}
I_1= 2\lambda^2+\lambda^{-4},
\qquad
I_2=2\lambda^{-2}+\lambda^{4},
\qquad
I_4=D_R^2,
\qquad
I_5=\lambda^{-4}D_R^2,
\qquad
I_6=\lambda^{-8}D_R^2.
\end{equation}

By using Eq.~\eqref{eq:PE_W}$_2$, we get
\begin{equation}\label{Er-expansion}
E_r  = 2\left(\lambda^2W^d_4+W^d_5+\lambda^{-6}W^d_6\right)D_R,
\end{equation}
where $W^s_j = \partial W^s/\partial I_j$, with $s=(d,\,e$).

The only non-zero components of the Cauchy stress in spherical coordinates are given by
\begin{align}\label{stress-component-dielectric}
&T_{rr}^d=2\left[\lambda^{-4}W^d_1+2\lambda^{-2}W^d_2+\left(\lambda^{-4}W^d_5+2\lambda^{-8}W^d_6\right)D_R^2\right]-p^d,
\quad\notag\\
&T_{\theta\theta}^d=T_{\psi\psi}^d=2\left[\lambda^{2}W^d_1+\left(\lambda^{-2}+\lambda^4\right)W^d_2\right]-p^d,
\end{align}
in the DE layer, and
\begin{align}\label{stress-component-elastic}
&T_{rr}^e=2\left(\lambda^{-4}W^e_1+2\lambda^{-2}W^e_2\right)-p^e,
\quad\notag\\
&T_{\theta\theta}^e=T_{\psi\psi}^e=2\left[\lambda^{2}W^e_1+\left(\lambda^{-2}+\lambda^4\right)W^e_2\right]-p^e,
\end{align}
in the elastic layer. Here %
$p^s$, where $s=(d,\,e)$, is a Lagrange multiplier associated with the incompressibility constraint of the $s$-th layer, which will be determined from the equilibrium equations and boundary conditions as detailed in the following.

The balance Eq.~\eqref{eq:balance} in the actual configuration imposes
\begin{equation}
    \label{eq:balance_cauchy}
\diver\tens{T}=\vect{0}.
\end{equation}

Under the symmetry assumption made in Eq.~\eqref{eq:symmetry}, Eq.~\eqref{eq:maxwell}$_2$ becomes
\begin{equation}
\label{eq:max1sphere}
\frac{1}{r^2}\frac{\partial (r^2D_r)}{\partial r}=0,
\end{equation}
which implies that $r^2D_r$ is a constant, and 
Eq.~\eqref{eq:balance_cauchy} reads
\begin{equation}\label{balance}
\frac{\partial T^s_{rr}}{\partial r}=\frac{2}{r}\left(T^s_{\theta\theta}-T^s_{rr}\right) \quad (s=d, e).
\end{equation} 

By introducing the reduced energy functions defined by
\begin{equation}\label{W-energy}
W_\text{sph}^d\left(\lambda,\,D_R\right)=W^d\left(I_1,\,I_2,\,I_4,\,I_5,\,I_6\right), \quad W_\text{sph}^e\left(\lambda\right)=W^e\left(I_1,\,I_2\right),
\end{equation}
and by substituting Eqs.~\eqref{invariants}, \eqref{Er-expansion} and \eqref{W-energy}, we have
\begin{equation}
E_r=\lambda^2\frac{\partial W_\text{sph}^d}{\partial D_R}.
\end{equation}

\noindent Accordingly,  the applied voltage is
\begin{equation}\label{connection}
V=\int_{r_i}^{r_m}\lambda^2\frac{\partial W_\text{sph}^d}{\partial D_R}\,\text dr.
\end{equation}
Similarly, Eq.~\eqref{balance} can be rewritten compactly as
\begin{equation}\label{reduced-balance}
\frac{\partial T_{rr}^s}{\partial r}=\frac{\lambda}{r}\frac{\partial W_\text{sph}^s}{\partial \lambda} \quad (s=d, \ e).
\end{equation}

Using the boundary conditions
\begin{equation}\label{static-boundary}
T^d_{rr}(r_i)=-P, \quad T^e_{rr}(r_o)=0,
\end{equation}
and Eq.~\eqref{reduced-balance}, the principal stresses in the radial direction read
\begin{align} \label{stress}
&{T^d _{rr}}= \int_{{\lambda_i}}^{{\lambda}} \frac{1}{1-\lambda^3}\frac{\partial W_\text{sph}^d}{\partial \lambda} {\mathrm{d}}\lambda-P \qquad &\text{at the DE layer}, \notag\\
&{T^e _{rr}}= -\int_{{\lambda_o}}^{{\lambda}} \frac{1}{1-\lambda^3}\frac{\partial W_\text{sph}^e}{\partial \lambda} {\mathrm{d}}\lambda \qquad &\text{at the elastic layer}.
\end{align}
Note that we have used the following relationship
\begin{equation}
\frac{{\mathrm{d} }r}{r}=\frac{\mathrm{d} \lambda}{\lambda \left(1-\lambda^3\right)},
\end{equation}
which can be obtained from Eqs.~\eqref{deformation_equation} and \eqref{stretch}.

The hoop stresses $T_{\theta\theta}^s=T_{\psi\psi}^s$ can be determined using the following relationship
\begin{equation}\label{circumferential-stress}
2(T_{\theta\theta}^s-T_{rr}^s)=\lambda\frac{\partial W_\text{sph}^s}{\partial\lambda}\qquad (s=d,e),
\end{equation}
which results from Eqs.~\eqref{balance} and \eqref{reduced-balance}.

The two layers are perfectly bonded to each other, imposing the continuity of the displacements and normal stresses at the interface, see Eq.~\eqref{eq:interface}. As a result, the relationship between $\lambda_i$, $\lambda_m$, $\lambda_o$ and $V$ can be established, using Eq~\eqref{stress}, as
\begin{equation} \label{boundary-condition}
\int_{{\lambda_i}}^{{\lambda_m}} \frac{1}{1-\lambda^3}\frac{\partial W_\text{sph}^d}{\partial \lambda} {\mathrm{d}}\lambda-P= -\int_{{\lambda_o}}^{{\lambda_m}} \frac{1}{1-\lambda^3}\frac{\partial W_\text{sph}^e}{\partial \lambda} {\mathrm{d}}\lambda.
\end{equation}

Finally, the deformation $\lambda_i$, $\lambda_m$ and $\lambda_o$ can be fully determined from Eqs.~\eqref{stretch-relationship}, \eqref{connection} and \eqref{boundary-condition}, once the inner pressure $P$, voltage $V$ and constitutive laws $W^s$, with $s=(d,\ e)$, are given.

\section{Linear stability analysis}
\label{section-incremental}

Mechanical instabilities may occur in finitely deformed solids, and the onset of buckling and necking can be predicted by using the theory of incremental deformations superposed on a finite strain \cite{hutchinson1974bifurcation, hill1975bifurcation, Zurlo_2017,Mao2019, Bortot18}. In this section, we derive the governing equations for the analysis of small-amplitude wrinkles superimposed upon the finite deformation of the balloon, and develop the surface impedance matrix method applicable to layered structures to build a robust numerical procedure for solving the resulting dispersion equations. Here we just give the general governing equations of the deformed DE elastomer, and  we omit the superscript for the sake of notation compactness. Note that the incremental governing equations for the elastic elastomer can be simply derived  by making the electric field vanish in the given solution.

\subsection{Incremental BVP}
Due to the spherical symmetry, the study of axisymmetric modes is sufficient to give a full insight into the buckling behavior of the material, since the dependence on $\psi$ does not alter the incremental BVP \cite{Haughton1978}. Let us superimpose a small axisymmetric incremental displacement $\dot{\vect{u}}=\dot{u}_r(r,\,\theta)\vect{ e}_r+\dot{u}_\theta (r,\,\theta)\vect{e}_\theta$ along with an incremental electric displacement $\dot{\vec D}=\dot D_r(r,\theta){\vec e}_r+\dot D_\theta (r,\theta) {\vec e}_\theta$ over the radially symmetric solution described in Section \ref{section-nonlinear}. Hereinafter the incremental quantity will be denoted by the notation ${\left(\dot{ \bullet  }\right)}$.

In the following, we adopt the convention of summation over repeated indices. The linearized incremental forms of the constitutive relations read \cite{Dorfmann2010}
\begin{equation}\label{incremental-constitive}
\begin{aligned}
&\dot{\tens{S}}=\mathcal{A}:\tens{H} +\boldsymbol{\tens{\Gamma}}\dot{\vect{D}}+p{\tens H}-\dot{p}{\tens I}, 
&&\dot{\vec E}=\tens{H}:\boldsymbol{\tens{\Gamma}}+\tens{K}\dot{\vect{D}},\\
&\dot{S}_{ij} = A_{ij\alpha\beta}H_{\beta\alpha} + \Gamma_{ij\alpha}\dot{D}_\alpha + p H_{ij} -\dot{p}\delta_{ij}, &&\dot{E}_i=H_{\alpha\beta}\Gamma_{\beta\alpha i}+K_{i\alpha}\dot{D}_\alpha.
\end{aligned}
\end{equation}
where $\tens{H}=\grad\dot{\vect{u}}$ is the displacement gradient, $\mathcal{A},\,\boldsymbol{\tens{\Gamma}}$ and $\tens{K}$ are, fourth-, third- and second-order electro-elastic moduli tensors, respectively, whose components are given by
\begin{equation}
\begin{gathered}
A_{piqj}=A_{qjpi}=F_{p\alpha}F_{q\beta}\frac{\partial^2W}{\partial F_{i\alpha}\partial F_{j\beta}},
\quad
\Gamma_{piq}=\Gamma_{ipq}=F_{p\alpha}F_{\beta q}^{-1}\frac{\partial^2W}{\partial F_{i\alpha}\partial D_{l\beta}},\\
K_{ij}=K_{ji}=F_{\alpha i}^{-1}F_{\beta j}^{-1}\frac{\partial^2W}{\partial D_{l\alpha}\partial D_{l\beta}}.
\end{gathered}
\end{equation}
The incremental counterpart of the equilibrium Eqs.~\eqref{eq:maxwell}-\eqref{eq:balance} read
\begin{align}
\label{eq:incremental_equilibrium1}
\diver\dot{\tens{S}}&=\vect{0},\\
\label{eq:incremental_equilibrium2}
\curl\dot{\vect{E}}&=\vect{0},\\
\label{eq:incremental_equilibrium3}
\diver \dot{\vect{D}}&=0.
\end{align}
We introduce an incremental electric potential $\dot\phi = \dot{\phi}(r,\,\theta)$ to rewrite the incremental electric field as $\dot{\vec E}=-\grad\dot\phi$, so that Eq.~\eqref{eq:incremental_equilibrium2} is automatically satisfied.

For the considered deformation, we have
\begin{equation} 
\tens{H}=\grad\dot{\vect{u}} = \left[ {\begin{array}{*{20}{c}}
	{\dfrac{{\partial {\dot{u}_r}}}{{\partial r}}}&{\dfrac{1}{r}\left( {\dfrac{{\partial {\dot{u}_r}}}{{\partial \theta }} - {\dot{u}_\theta }} \right)}&0 \vspace{2ex} \\
	{\dfrac{{\partial {\dot{u}_\theta }}}{{\partial r}}}&{\dfrac{1}{r}\left( {{\dot{u}_r} + \dfrac{{\partial {\dot{u}_\theta }}}{{\partial \theta }}} \right)}&0 \vspace{2ex} \\
	0&0&{\dfrac{1}{r}\left( {{\dot{u}_r} + {\dot{u}_\theta }\cot \theta } \right)}
	\end{array}} \right].
\end{equation}

The incompressibility Eq.~\eqref{eq:inc} at the incremental order reads
\begin{equation}\label{incremental-incompressibility1}
\diver\dot{\vect{u}}=0.
\end{equation}

Having assumed that the applied voltage and pressure are fixed and that the two layers are perfectly bonded, the following boundary and interfacial conditions apply
\begin{align}\label{incremental-boundary-components}
&\dot{S}^d_{rr}=P\frac{\partial \dot{u}_r}{\partial r}, \quad \dot{S}^d_{r\theta}=\frac{P}{r}\left(\frac{\partial \dot{u}_r}{\partial \theta}-\dot{u}_\theta\right), \quad \dot{\phi}=0 &\text{at}&\quad r=r_i,\\ \notag
&\dot{S}_{rr}^d=\dot{S}_{rr}^e, \quad \dot{S}^d_{r\theta}=\dot{S}^e_{r\theta}, \quad \dot{\phi}=0  &\text{at}&\quad r=r_m,\\ \notag
&\dot{S}^e_{rr}=\dot{S}^e_{r\theta}=0 &\text{at}&\quad r=r_o.
\end{align}

\subsection{Stroh formulation}
We assume the following separation of variables for the incremental fields \cite{martin2020stroh}
\begin{equation} \label{incremental-solution}
\begin{aligned}
&\left\{ {{\dot{u}_r}\left( {r,\theta } \right),{{\dot S}_{rr}}\left( {r,\theta } \right),\dot \phi \left( {r,\theta } \right),{{\dot D}_{r}}\left( {r,\theta } \right)} \right\} = \left\{ {{U_r}\left( r \right),{\Sigma _{rr}}\left( r \right),\Phi \left( r \right),{\Delta _r}\left( r \right)} \right\}{P_m}\left( {\cos \theta } \right),\\
&\left\{ {{\dot{u}_\theta }\left( {r,\theta } \right),{{\dot S}_{r\theta }}\left( {r,\theta } \right)} \right\} = \left\{ {\frac{{{U_\theta }\left( r \right)}}{{M }},\frac{{{\Sigma_{r\theta }}\left( r \right)}}{{M }}} \right\}\frac{\d P_m\left(\cos \theta\right)}{\d\theta},
\end{aligned}
\end{equation}
where $M=\sqrt {m\left( {m + 1} \right)}$, and $P_m$ indicates the Legendre polynomial of order $m$, which satisfies the following identity
\begin{equation} 
\frac{\d^2 P_m \left( {\cos \theta } \right)}{\d\theta ^2} + \cot \theta \frac{\d P_m \left( {\cos \theta } \right)}{\d\theta } + M^2{P_m}\left( {\cos \theta } \right) = 0.
\end{equation}

Given the Stroh vector $\vect{\eta}(r)=\left(U_r,\,U_\theta,\,r\Delta_r,\,r\Sigma_{rr},\,r\Sigma_{r\theta},\, \Phi 
\right)$, the governing equations \eqref{incremental-constitive}, \eqref{eq:incremental_equilibrium1}, \eqref{eq:incremental_equilibrium3} and \eqref{incremental-incompressibility1} can be rewritten in the form of a first-order differential system as 
\begin{equation} \label{stroh-formulation}
\frac{\text d}{\text dr}\vect{\eta} = \frac{1}{r}\tens{G}\vect{\eta} = \frac{1}{r}\begin{bmatrix}
	\tens{G}_1 &\tens{G}_2\\
	\tens{G}_3 &\tens{G}_4
	\end{bmatrix} {\vec{\eta }},
\end{equation} 
where the matrix $\tens{G}\in \mathbb{R}^{6\times 6}$ is the so-called Stroh matrix. The derivation of Eq.~\eqref{stroh-formulation} and the components of the 3 $\times$ 3 sub-matrices $\tens{G}_1,\,\tens{G}_2,\,\tens{G}_3$ and $\tens{G}_4$ are detailed in Supplementary Material.

We introduce the generalized displacement and traction vectors, defined as $\vect{U}=\left(U_r,\,U_\theta,\,r\Delta_r\right)$ and $\vect{S}=\left(r\Sigma_{rr},\,r\Sigma_{r\theta},\,\Phi 
\right)$, respectively. Then, using Eq.~\eqref{incremental-solution}, the incremental boundary conditions \eqref{incremental-boundary-components}$_{1, 3}$ can be rewritten as
\begin{equation} \label{incremental-boundary-stroh}
{\vec{S}^d}\left( {{r_i}} \right) = {P}\begin{bmatrix}
	{ - 2}&{M }&0\\
	{M }&{ - 1}&0\\
	0&0&0
	\end{bmatrix}{\vec{U}^d}\left( {{r_i}} \right), \quad {\vec{S}^e}\left( {{r_o}} \right) = {\vec{0}},
\end{equation}
and the incremental interfacial condition \eqref{incremental-boundary-components}$_{2}$ can be rewritten as
\begin{equation} \label{incremental-interfacial-stroh}
{\vec{S}^d}\left( {{r_m}} \right) ={\vec{S}^e}\left( {{r_m}} \right) , \quad {\vec{U}^d}\left( {{r_m}} \right) = {\vec{U}^e}\left( {{r_m}} \right).
\end{equation}

\subsection{The surface impedance matrix method}
Here we exploit the so-called surface impedance matrix method to build a robust numerical procedure to solve the governing equation \eqref{stroh-formulation} associated with the incremental boundary condition \eqref{incremental-boundary-stroh} and interfacial condition \eqref{incremental-interfacial-stroh}.

For each layer, we introduce the conditional impedance matrices  $\tens{Z}^d(r,\,r_i)$ and $\tens{Z}^e(r,\,r_o)$ \cite{Norris_2010}. In particular, we have 
\begin{equation} \label{32}
\vect{S}^s = \tens{Z}^s\vect{U}^s\qquad (s=d,\,e).
\end{equation}
Then we can expand Eq.~\eqref{stroh-formulation} to obtain (with $s$ omitted)
\begin{equation}\label{two_Riccati}
\frac{\d}{\d r}\vec U=\frac{1}{r}\tens{G}_1 \vec U+\frac{1}{r}\tens{G}_2 \tens{Z} \vect{U}, \quad
\frac{\d}{\d r}(\tens{Z} \vect{U})=\frac{1}{r}\tens{G}_3 \vect{U}+\frac{1}{r}\tens{G}_4 \tens{Z} \vect{U}.
\end{equation}
Elimination of $\vect{U}$ in Eq.~\eqref{two_Riccati} gives the following Riccati differential equation for ${{\tens{Z}}}$
\begin{equation} \label{Riccati}
\frac{\d \tens{Z}}{\d r} = \frac{1}{r}\left(  - \tens{Z}\tens{G}_1 - \tens{Z}\tens{G}_2\tens{Z}+ \tens{G}_3 + \tens{G}_4\tens{Z}\right).
\end{equation}
From the incremental boundary condition \eqref{incremental-boundary-stroh}, we have
\begin{equation} \label{initial-target-condition}
\begin{array}{l}
\tens{Z}^d\left( r_i,\,r_i\right) = P\left[ {\begin{array}{*{20}{c}}
	{ - 2}&{M }&0\\
	{M }&{ - 1}&0\\
	0&0&0
	\end{array}} \right],\\
{\tens{Z}^e\left(r_o,\,r_o \right) = \tens{0}}.
\end{array}
\end{equation}

The marginal stability curves for the layered balloon can be determined as follows. First, we determine the deformation and material constants  for a given voltage $V$ and a given inner pressure $P$, based on the results presented in Section \ref{section-nonlinear}. Then, we integrate Eq.~\eqref{Riccati}  in the DE layer from $r_i$ to $r_m$, in order to obtain $\tens{Z}^d(r_m,\,r_i)$, with the initial condition \eqref{initial-target-condition}$_1$. In the elastic layer, we integrate Eq.~\eqref{Riccati} from $r_o$ to $r_m$ to obtain $\tens{Z}^e(r_m,\,r_o)$, with the initial condition \eqref{initial-target-condition}$_2$. We finally iterate on the stretch until the following bifurcation criterion is satisfied
\begin{equation}\label{bifurcation-equation}
{\det\left[ \tens{Z}^e(r_m,\,r_o)-\tens{Z}^d(r_m,r_i)\right]=0.}
\end{equation}

For the considered problem, the critical inner stretch for the onset of a mechanical instability can be solved from the dispersion equation \eqref{bifurcation-equation}, which is a function of the applied voltage, the pressure, the mode $m$, and the material and structural parameters of the balloon, such that:
\begin{equation}
\lambda_i^c=\lambda_i^c(V,P,m;\mu^d,\mu^e,\varepsilon,t^d,t^e).
\end{equation}

\section{Marginal stability curves for Gent dielectric-elastomer balloons}
\subsection{Constitutive Equations}
\label{section-energy}
In order to illustrate the results of the linear stability analysis,  we adopt the following ideal \emph{Gent dielectric model} \cite{Li2013} and the \emph{Gent elastic model} \cite{Gent99} to describe the DE and elastic elastomers, respectively,
\begin{align}\label{energy}
&W^d(\tens{F},\,\vect{D}_R)=-\frac{\mu^dG^d}{2}\ln \left(1-\frac{I_1-3}{G^d}\right)+\frac{I_5}{2\varepsilon},\\ \notag
&W^e(\tens{F})=-\frac{\mu^eG^e}{2}\ln \left(1-\frac{I_1-3}{G^e}\right),
\end{align}
where $\mu^s$ and $G^s$ are the shear modulus and the dimensionless stiffening parameter of the $s$-th $(s=d, e)$ elastomer, respectively, $\varepsilon$ is the strain independent permittivity of the DE elastomer.
These constitutive laws model the strain stiffening behavior of soft polymers. In the following, we fix $G^d=G^e=97.2$, an experimental parameter collected for unfilled vulcanized rubber  \cite{Gent99, Dorfmann2014}. Note that in the limit of $G^s\to \infty$, the Gent model reduces to the neo-Hookean model \cite{Zhao2007, Kim2012}.

\subsection{Marginal stability curves}
\label{section-numerical}
We now derive the marginal stability curves to investigate the influence of the presence of the elastic inactive layer on the deformation and instabilities of the DE active balloon, and explore the possibility of realizing the selection of specific instability mode in layered DE devices through structural and material design.
For this purpose, we compare the nonlinear responses, as well as the onsets of the snap-through and the buckling  instabilities of a monolayer elastic balloon, a monolayer DE balloon and a layered dielectric-elastic balloon.

For convenience and generality, we introduce the following dimensionless quantities,
\begin{align}
&\overline P=\frac{P}{\mu^d},& \quad &\overline{T}_{ii}^s=\frac{T_{ii}^s}{\mu^d}\ \ \ (s=d,e),\\ \notag 
&\overline V=\frac{V}{R_m-R_i}\sqrt{\frac{\varepsilon}{\mu^d}}, & \quad &\overline D_R=\frac{D_R}{\sqrt{\mu^d\varepsilon}}.
\end{align}

\subsubsection{Results for a monolayer elastic balloon}
\begin{figure}[t!]
	\centering
	\includegraphics[width=1\textwidth]{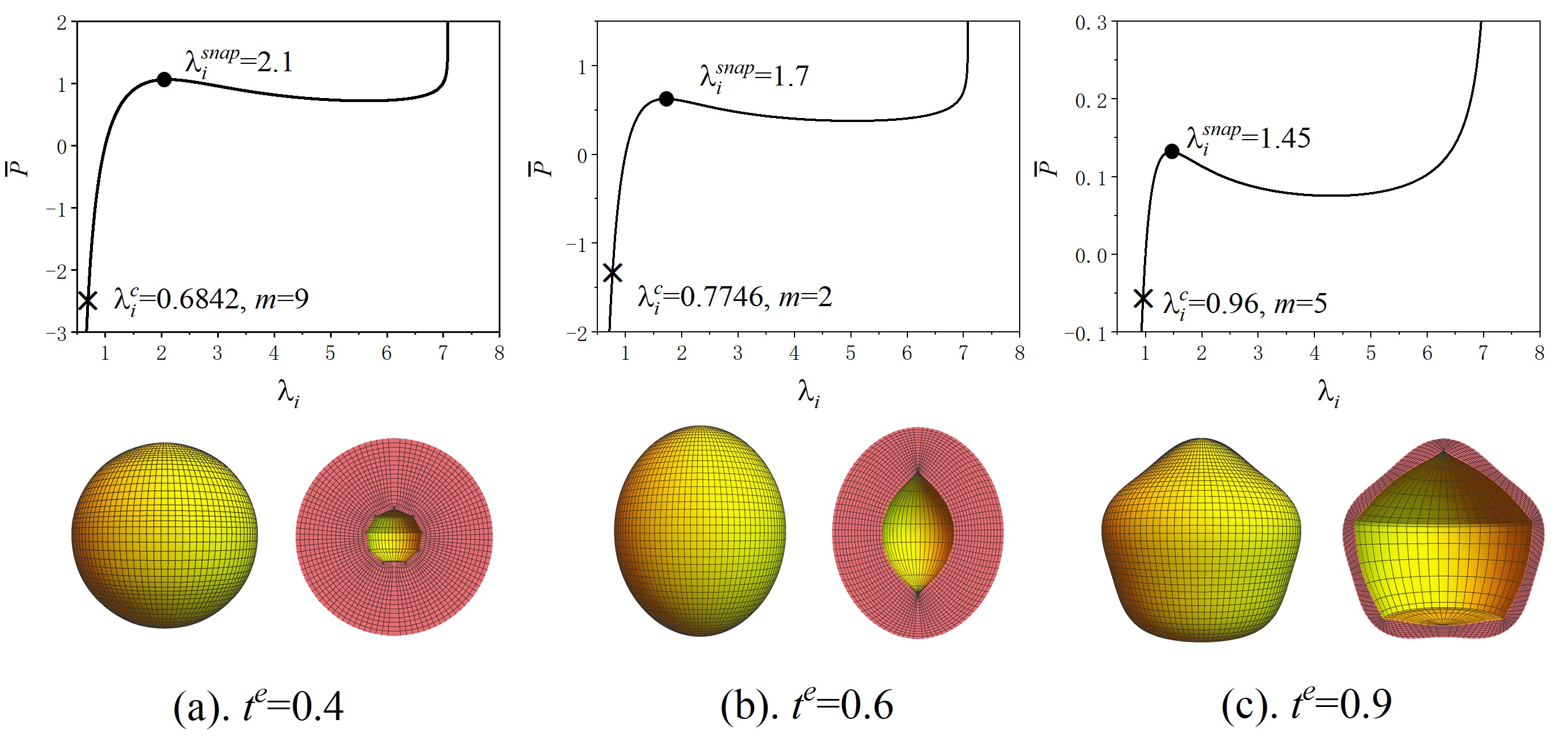}
	\caption{
			Inflation and buckling responses of (a) thick, (b) moderately thick and (c) thin elastic balloons ($\overline V=0,\ t^d=1$), respectively. The cross $\times$ and circle $\bullet$ markers indicate the thresholds for buckling in compression and snap-through instabilities of the material, respectively. The upper row presents the $\overline P-\lambda_i$ curve, and the lower row shows the shape of the balloon at the onset of buckling. Note that we have set a finite amplitude of the incremental displacement for the sake of graphical illustration of the buckling pattern.
	}
	\label{figure3}
\end{figure}

We first consider the case of an elastic balloon ($\overline V=0,\ t^d=1$) subject to an internal pressure $\overline P$, whose results are depicted in Fig. \ref{figure3}. We can see that the $\overline P-\lambda_i$ curve of the balloon is clearly non-monotonic and the snap-through behavior can be observed at the critical stretches highlighted by the round markers. The stretch $\lambda_i$ first increases as the pressure $\overline P\ (>0)$ increases. Once the pressure reaches a critical value, the stretch increase suddenly and the pressure inside the balloon decreases due to the dramatic increase of the volume. Due to the strain-stiffening effect of the material, the internal pressure increases again as the stretch approaches the extensible limit of the material. The snap-through enables a large strain change in the balloon, which is a desired actuation mechanism in many engineering applications. It is noted that the critical internal pressure for triggering the snap-through instability of a thin balloon is smaller than that of a thick balloon.

On the other hand, buckling  may occur in an elastic balloon subject to critical compression ($\overline P<0$). Compared with a thick-walled balloon, a balloon with a smaller thickness is more susceptible to buckling. We note that the critical mode $m$ (i.e. the first mode to become unstable as the pressure decreases) can be selected by properly designing the thickness of the balloon. Typically, the outer contour of a buckled thick-walled balloon maintains the spherical configuration and wrinkles appear on the inner face (Fig. \ref{figure3}a). While for a thin-walled balloon, buckling affects the whole body (Figs. \ref{figure3}$b,\ \ref{figure3}c$). 

\begin{figure}[t!]
	\centering
	\includegraphics[width=0.45\textwidth]{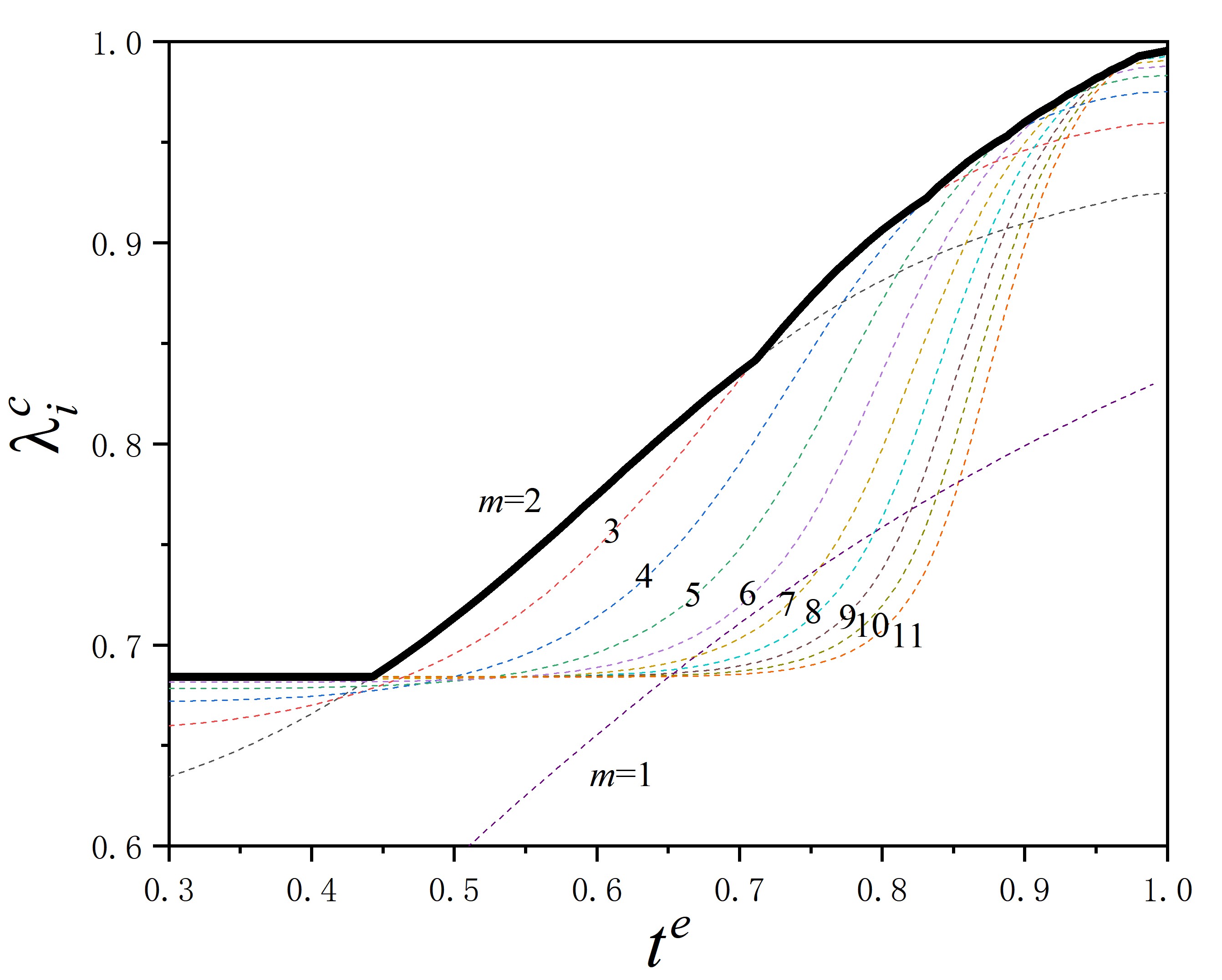}
	\caption{
		Plots of the critical strains  of a Gent elastic balloon as functions of the radius ratio $t^e$, shown at  different buckling modes $m=1,2,\dots,11$ (dashed lines). The bold black line corresponds to the marginal stability curve.
	}
	\label{figure4}
\end{figure}

In Fig.~\ref{figure4},  we plot the critical stretch  versus the aspect ratio $t^e$ at different buckling modes, computed from the dispersion equation \eqref{bifurcation-equation}. Buckling occurs once the stretch reaches the marginal stability threshold, i.e. the bold black curve in the figure. We note that the buckling mode $m=1$ is not allowed independently of the thickness of the balloon. The buckling mode $m=2$ can occur in moderately thick balloons ($0.45<t^e<0.7$), while for thick balloons ($t^e\leq 0.45$) and thin balloons ($t^e\geq0.75$),  higher mode ($m>2$) are selected, strongly depending on the aspect ratio. We remark that for thick balloons with $t^e \leq 0.45$, buckling always occurs once the inner circumferential stretch reaches $\lambda_i^c$=0.684, which is thickness independent.

We emphasize that a critical compression ($\overline P<0$) is required  for a Gent elastic balloon to trigger the onset of  buckling, which is different from the DE case presented below.

\subsubsection{Results for a monolayer DE balloon}
\begin{figure}[t!]
	\centering
	\includegraphics[width=0.75\textwidth]{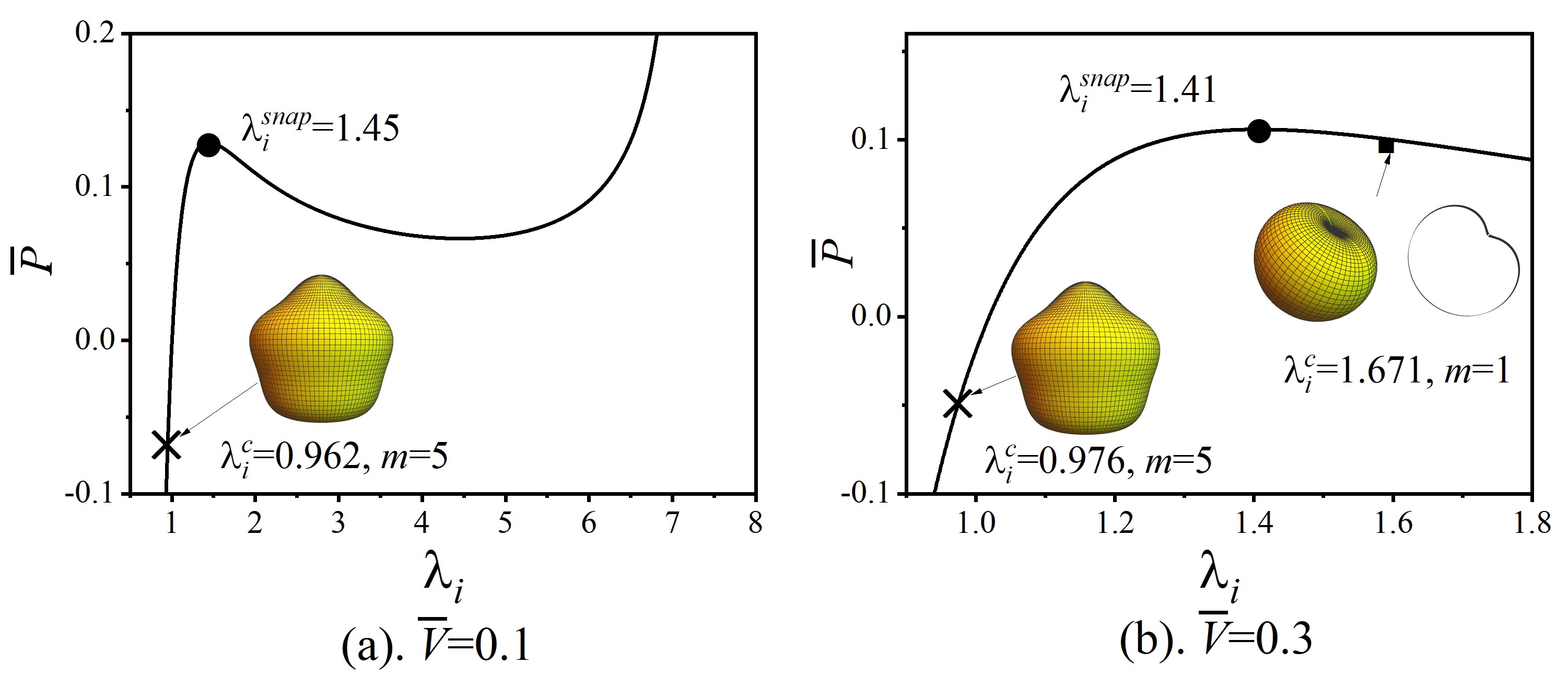}
	\caption{
			Inflation and buckling responses of thin DE balloons ($t^e=1,\ t^d=0.9$) subject to a fixed voltage: (a) $\overline V=0.1$ and (b) $\overline V=0.3$. The cross $\times$, circle $\bullet$ and square $\blacksquare$  markers indicate the thresholds for buckling in compression, snap-through and buckling in extension instabilities of the material, respectively. Inset: the shape of the balloon at the onset of buckling; and the sectional view of the balloon at necking.
	}
	\label{figure5}
\end{figure}

Here, we consider a thin DE balloon ($t^e=1, t^d=0.9$) subject to a combination of an internal pressure $\overline P$ and a voltage $\overline V$, whose results are collected in Fig. \ref{figure5}. In order to investigate the effect of the applied voltage on the onset of buckling, we consider the cases of $\overline V=0.1$ and $\overline V=0.3$ as illustrative  examples.

For a DE balloon subject to a small voltage ($\overline V=0.1$), the response is similar to the case of a purely elastic monolayer presented in Fig. \ref{figure3}. The balloon buckles only when a critical compression ($\overline P<0$) is applied. On the other hand, the snap-through instability is triggered once a critical inner pressure $\overline P$ ($>0$) is applied, and the balloon survives the snap-through instability, without encountering the buckling failure.

When the applied voltage is sufficiently large ($\overline V=0.3$), we see that in addition to buckling in compression, the possibility of a bifurcation in extension ($\overline P>0$)  emerges along the path of the snap-through. In this case, the balloon can not reach a homogeneous state characterized by a large strain. 

\begin{figure}[t!]
	\centering
	\includegraphics[width=0.9\textwidth]{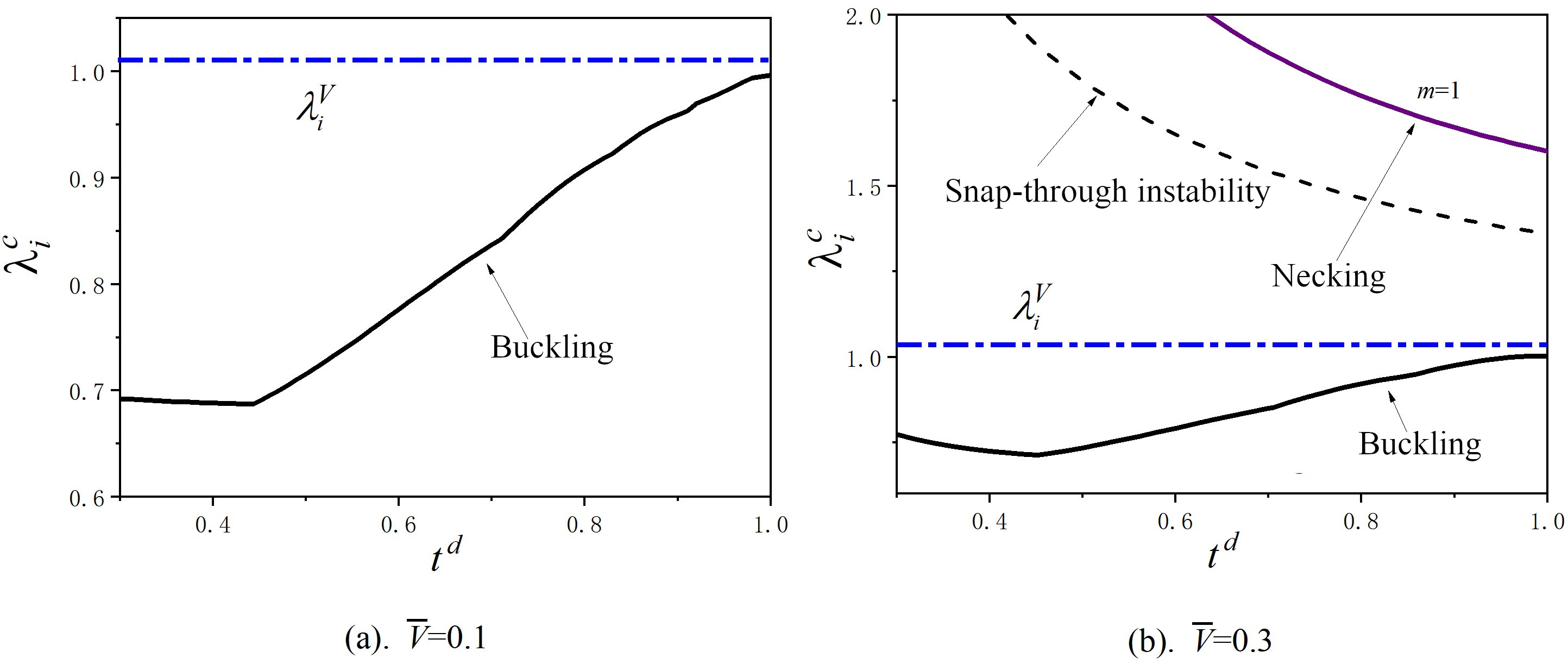}
	\caption{
		{
			Plots of the critical strains versus the radius ratio $t^d$ for a single layer of DE ($t^e=1$): (a) $\overline V=0.1$ and (b) $\overline V=0.3$.  The black dashed line depicts the snap-through instability threshold, the blue dot-dashed line depicts the inner circumferential stretch $\lambda_i^V$ of the balloon induced by the applied voltage only, i.e. considering $\overline P=0$. The magenta solid line finally depicts the threshold for necking. For simplification, here we do not show the instability thresholds for the different modes $m=1,\,2,\dots,11$ (dashed lines as in Fig. \ref{figure4}), but only present the actual marginal stability curve. 
		}
	}
	\label{figure6}
\end{figure}

We plot in Fig. \ref{figure6} the critical stretch $\lambda_i^c$ versus $t^d$ for monolayer DE balloons subject to either $\overline V=0.1$ or $\overline V=0.3$. Here we use the notation $\lambda_i^V$ to denote the inner circumferential stretch of the balloon induced by the applied voltage only, i.e. considering $\overline P=0$. 

For the case $\overline V=0.1$, the balloon expands radially to $\lambda_i^V=1.014$, until buckling occurs at a critical compression. The snap-through instability of the elastomer will be triggered  at a critical inner pressure $\overline P>0$ (See Fig. \ref{figure5} (left) for the special case $t^d=0.9$). We note that no bifurcation occurs during the snap-through process, thus the balloon can achieve a large actuation strain (In Fig. \ref{figure6} (left) the snap-through curve is not presented). 

For the case $\overline V=0.3$, in addition to the buckling in compression, a bifurcation in tension also occurs. The snap-through  occurs prior to the bifurcation, thus  snap-through cannot be exploited  to obtain a large actuation strain, since the balloon would lose its spherical configuration during the snap-through process. As shown in Fig. \ref{figure5} (right), the critical mode for the bifurcation in tension is $m=1$, and a localized thinning of the DE occurs, which is the typical feature of necking instability \cite{hutchinson1974bifurcation, hill1975bifurcation}. It is noted that we can design the buckling mode in compression  by properly selecting the thickness of the balloon, while $m=1$ is always the critical mode in tension.

We conclude that due to the electro-mechanical coupling effect, the DE balloon may undergo snap-through-induced necking, thus limiting its applicability as an actuating device.

\subsubsection{Results for a bi-layered dielectric-elastic balloon}

Finally,  we investigate the case of a bi-layered dielectric-elastic balloon, as both voltage and pressure tuning methods are considered.

\begin{figure}[t!]
	\centering
	\includegraphics[width=0.85\textwidth]{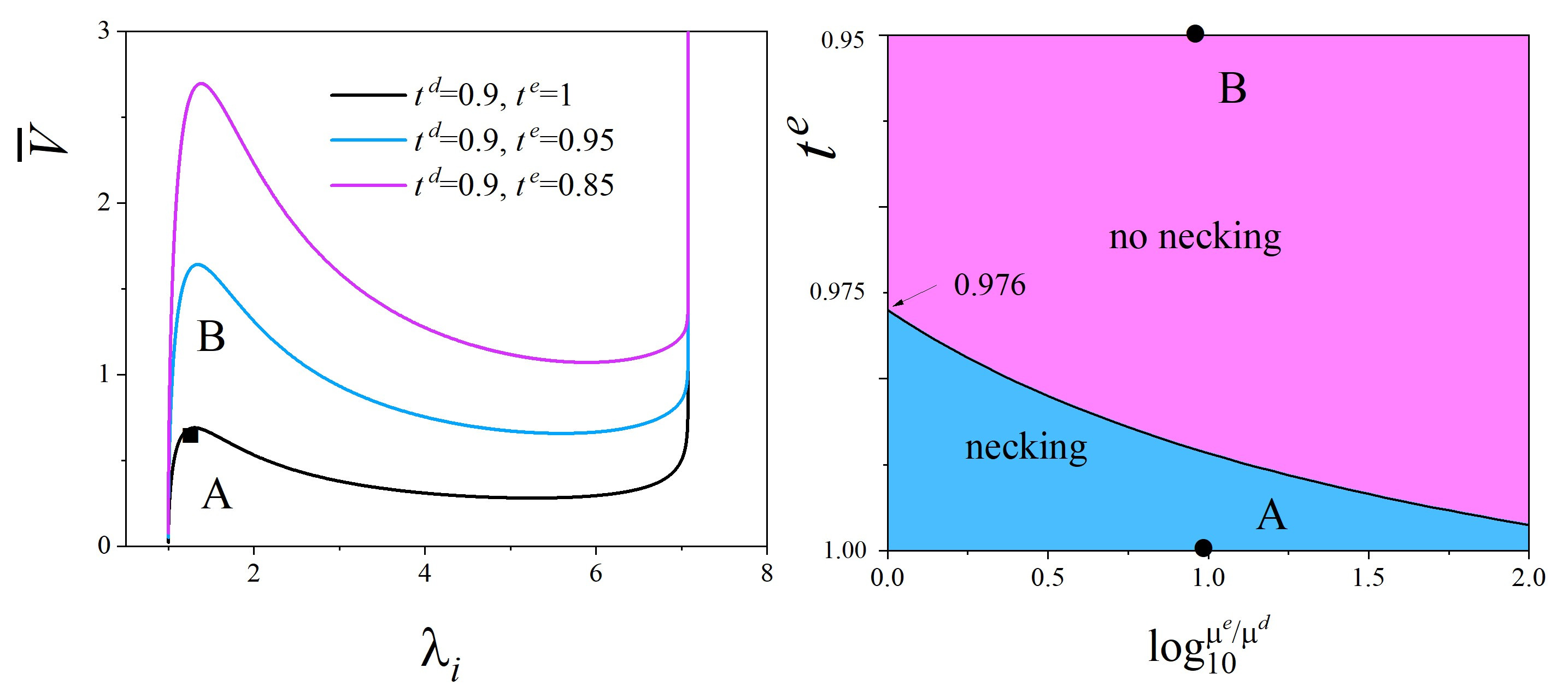}
	\includegraphics[width=0.85\textwidth]{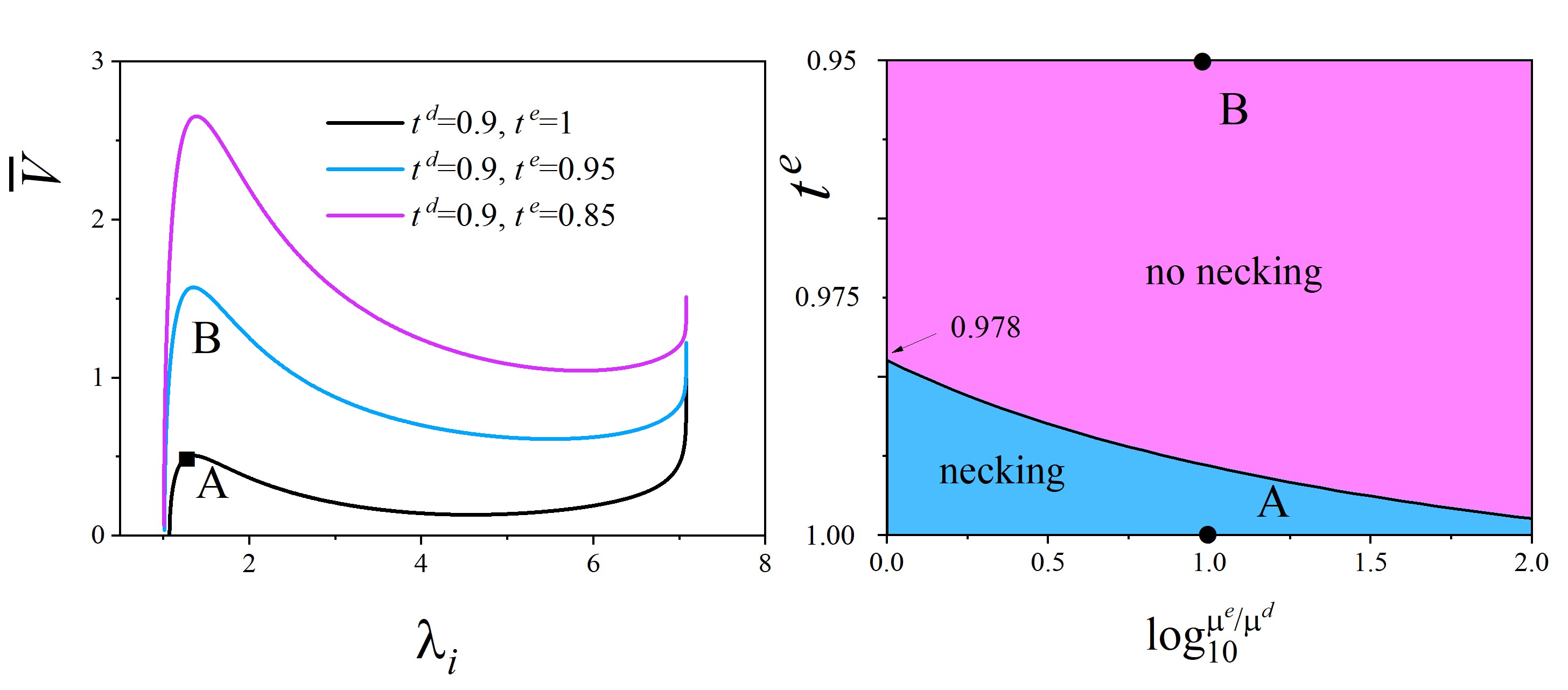}
	\caption{
			Inflation of voltage-activated ($\overline P=0$ top, $\overline{P}=0.06$ bottom) dielectric-elastic balloons with $t^d=0.9$ and varying $t^e$. Left:  $\overline V-\lambda_i$ curve of the balloon at fixed  $\mu^e/\mu^d=10$ and varying $t^e$, with the square $\blacksquare$ marker indicating the necking threshold. Right:  stability diagram  of the balloons with varying $t^e$ and $\text{log}^{\mu^e/\mu^d}_{10}$. The black, blue and red curves on the left column correspond to three specific balloons: balloon $A$ with $t^d=0.9,\ t^e=1$, balloon $B$ with $t^d=0.9,\ t^e=0.95$, and balloon $C$ with $t^d=0.9,\ t^e=0.85$ (balloon $C$ beyond the scope of the phase diagram thus is not presented in the plots on the right). 
	}
	\label{figure7}
\end{figure}

In Fig. \ref{figure7} (left) we show the voltage-stretch curves and the necking thresholds of electro-active bi-layered balloons ($\overline P=0$ and $0.06$) with $t^d=0.9$ and varying $t^e,\ \mu^d$ and $\mu^e$. We consider the balloons with fixed material parameter $\mu^e/\mu^d=10$. The results for a monolayer DE balloon are presented here for comparison. We remark that the snap-through instability always exists in bi-layered balloons, which is independent of the thickness of the elastic layer. Compared with balloons with a thin elastic layer, a larger voltage is required to activate the snap-through instability in balloons with a thick elastic layer, increasing the risk of electric breakdown failure. It is worth noting that for dielectric-elastic balloons with a thin elastic layer (here $t^e=1$ for example), a bifurcation in tension (i.e. necking instability) occurs after the snap-through instability is triggered, although the thresholds of the critical stretches for the two mechanical instabilities are very close. As expected, covering the DE balloon with an elastic layer with specific thickness (here $t^e=0.95,\ 0.85$ for example) can suppress the bifurcation in tension. As a result, the structure can survive the snap-through procedure and achieve large deformation without necking, thus being suitable for engineering applications as an actuator. We also notice that as the pressure increases, the critical voltage required to trigger the snap-through instability decreases. 

In order to study the influence of the thickness of the elastic layer on the necking instability of the bi-layered balloon, we plot the $t^e -$ $\log^{\mu^e/\mu^d}_{10}$ phase diagram on Fig. \ref{figure7} (right). For $\overline{P}=0$, we can see that the necking may occur when the thickness of the elastic layer is small. As the thickness of the elastic layer increases (e.g., $t^e<0.976$ for balloons with $\mu^d=\mu^e$), the necking of the balloon can be suppressed. As the internal pressure or the stiffness of the elastic layer ($\mu^e=10\mu^d$ for example) increase, we notice a decrease in the critical thickness of the elastic layer required to suppress the necking instability.

\begin{figure}[t!]
	\centering
	\includegraphics[width=0.65\textwidth]{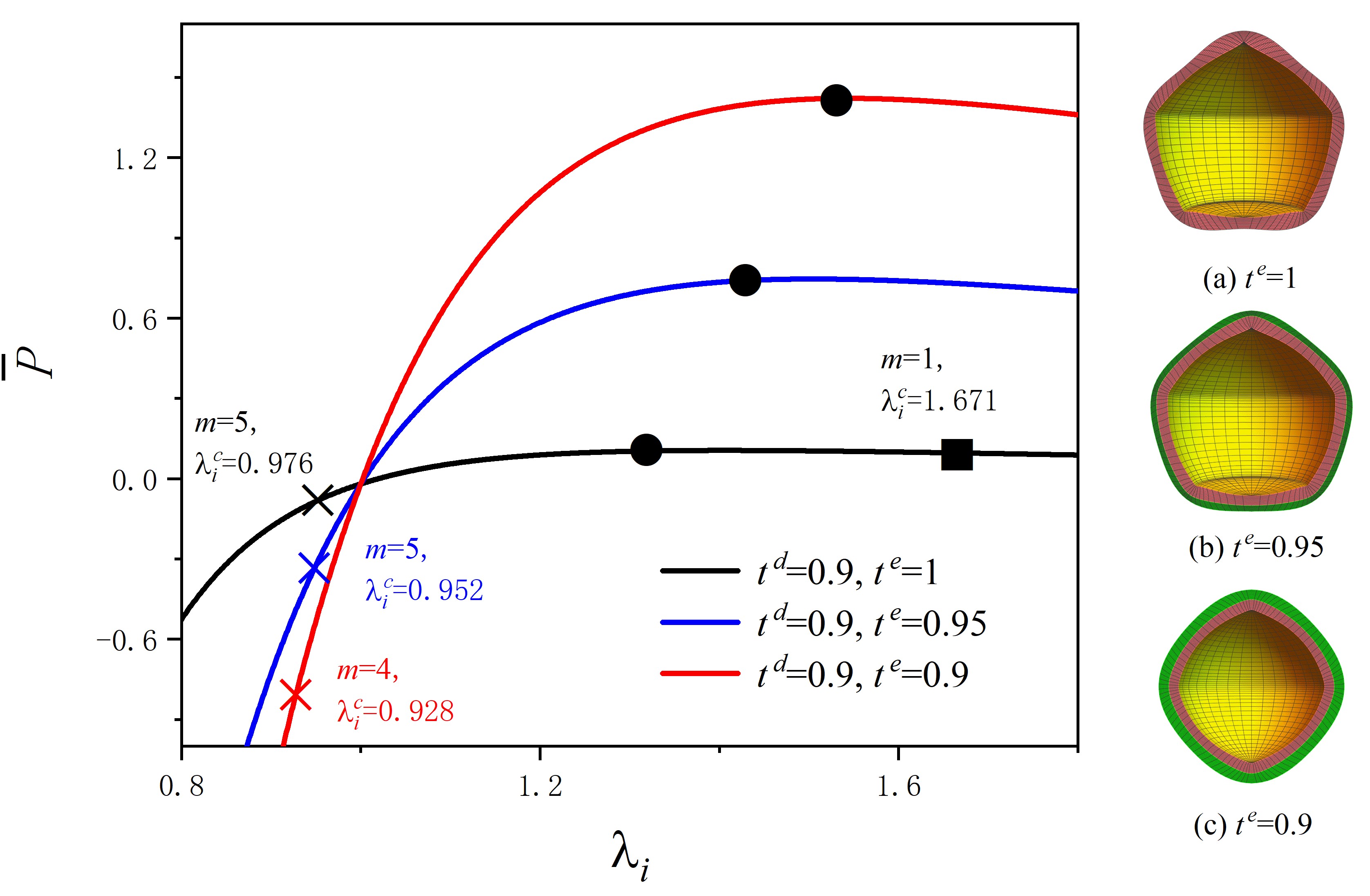}
	\caption{
		{\footnotesize
			Inflation and buckling responses of pressure activated ($\overline V=0.3$) dielectric-elastic balloons with $t^d=0.9$ and varying $t^e$. The  $\overline P-\lambda_i$ curve is shown on the left, and the buckling shapes are shown on the right. The cross $\times$, circle $\bullet$ and square $\blacksquare$ markers indicate the thresholds for buckling in compression, snap-through and buckling in extension, respectively.
		}
	}
	\label{figure8}
\end{figure}

Fig. \ref{figure8} presents the inflation-stretch curves bi-layered balloons ($\overline V=0.3$) with $t^d=0.9$, $\mu^e/\mu^d=10$ and varying $t^e$. We remark that for the case of a monolayer DE layer ($t^e=1$), both buckling and necking can be induced by properly tuning the pressure $\overline P$. The presence of the elastic layer can not only decrease the risk of buckling but can also suppress the necking of the balloon. The selection of buckling pattern of the balloons can be designed by setting the thickness of the elastic layer.

We conclude that for a monolayer DE balloon, the tensile necking may be triggered during the snap-through process, limiting the actuation strain. The application of the elastic layer outside the DE layer can enhance the snap-through instability to avoid a bifurcation in tension, thus can dramatically increase the actuation strain of the balloon. That is, a monolayer DE balloon is more suitable for making functional devices with complex surface morphology, while a bi-layered dielectric-elastic balloon is more suitable  as an actuating device  as stable large deformation can be achieved.

\section{Fully nonlinear numerical simulations}
\label{section-post-bifurcation}
In this Section, we first implement a numerical scheme to approximate the nonlinear BVP given by Eqs.~\eqref{eq:inc}-\eqref{eq:boundary2}. Second, we perform the numerical simulations, discussing the morphological transitions in the fully nonlinear regime in some cases of practical interest.

\subsection{Variational formulation}
We take  the displacement field $\vect{u}$ and the electric potential field $\phi$ as the unknowns of the problem. 
We assume radial symmetry of the solutions  performing numerical simulations of an axis-symmetric section of the balloon, namely on $\mathcal{S}_0 = \mathcal{B}_0\cap(X_3=0\text{ and } X_1>0)$.

For the sake of numerical robustness, we use a quasi-incompressible approximation.
The BVP given by Eqs.~\eqref{eq:inc}-\eqref{eq:boundary2} rewrites
\begin{equation}
\label{eq:nonlin_sist}
\sistema{
&\tens{S} = \frac{\partial W^\star}{\partial \tens{F}},\quad \vect{D}_L = -\frac{\partial W^\star}{\partial \vect{E}_L},\\
&\Diver \tens{S} = \vect{0} &&\text{in }\mathcal{S}_0\\
&\Diver \vect{D}_L = \vect{0} &&\text{in }\mathcal{S}_0^d\\
&\tens{S}^T \vect{N} = -P\tens{F}^{-T}\vect{N},\quad \phi = 0 &&\text{if }R=R_i\\ 
&\tens{S}^T\vect{N} = \vect{0} &&\text{if }R=R_o\\
&\phi = V &&\text{if }R\geq R_m\\
&\vect{u}\cdot\vect{e}_1 = 0, \quad\vect{e}_2\cdot\tens{S}^T\vect{e}_1 = 0 &&\text{if }X_1=0\\
&\int_{\mathcal{S}_0} \rho\vect{u}\cdot{\vect{e}_2}\,\d X_1 \d X_2 = 0
}
\end{equation}
where $(\vect{e}_1,\,\vect{e}_2,\,\vect{e}_3)$ is the canonical vector basis in Cartesian coordinate and $\rho = \sqrt{X_1^2+X_2^2}$. The last equation in \eqref{eq:nonlin_sist} removes rigid body translations along the vertical direction $\vect{e}_2$, while the energy density $W^\star(\tens{F},\,\vect{E}_L)$ is linked to $W(\tens{F},\,\vect{D}_L)$ through the following Legendre transform
\begin{equation}
W(\tens{F},\,\vect{D}_L) = W^\star(\tens{F},\,\vect{E}_L) + \vect{D}_L\cdot\vect{E}_L.
\end{equation}
The boundary value problem given by Eqs.~\eqref{eq:inc}-\eqref{eq:boundary2} can be cast into an equivalent variational formulation. 
The integral condition in \eqref{eq:nonlin_sist} is enforced by means of a Lagrange multiplier $\alpha$.

Let us introduce the energy functional
\begin{equation}
\mathcal{E}[\vect{u},\,\phi,\,\alpha] = 2\pi\int_{\mathcal{S}_0} \rho \left[W^\star(\tens{F},\,\vect{E}_L) + \alpha\vect{u}\cdot\vect{e}_2\right]\,\d X_1\d X_2.
\end{equation}
Then, the solutions of the BVP \eqref{eq:inc}-\eqref{eq:boundary2} must satisfy \cite{Bustamante_2008}
\begin{equation}
\delta \mathcal{E}(\vect{u},\,\phi,\,\alpha)[\delta\vect{u},\,\delta\phi,\,\delta\alpha]+2\pi\int_{R=R_i} J \rho P \tens{F}^{-1}\delta\vect{u}\cdot\vect{N} \,\d S = 0.
\end{equation}
where $J = \det\tens{F}$, $\delta \mathcal{E}$ is the first variation of the energy functional $\mathcal{E}$ and $\delta\vect{u},\,\delta\phi,\,\delta\alpha$ are admissible variations of the unknown fields.

As a compressible counterpart of Eq.~\eqref{energy}, we take the energy densities for the DE and the passive elastomer as follows
\begin{equation}
\left\{
\begin{aligned}
&{W^{\star}}^d(\tens{F},\,\vect{E}_L)=-\frac{\mu^d G^d}{2}\log\left(1-\frac{\overline{I}_1-3}{G^d}\right)+\frac{K^d}{2}\left(\log J\right)^2-\frac{\epsilon J}{2}\vect{E}_L\cdot\tens{C}^{-1}\vect{E}_L\\
&{W^{\star}}^e(\tens{F})=-\frac{\mu^e G^e}{2}\log\left(1-\frac{\overline{I}_1-3}{G^e}\right)+\frac{K^e}{2}\left(\log J\right)^2\\
\end{aligned}
\right.
\end{equation}
where $K^d$ and $K^e$ are parameters regulating the compressibility of each layer and $\overline{I}_1=J^{-2/3}I_1$.

\subsection{Mixed finite element implementation}
The problem is approximated by means of the finite element method.
The computational domain $\mathcal{S}_0$ is discretized by using a triangular mesh.
In order to impose homogeneous boundary conditions on the potential field, we decompose the potential $\phi$ into two contributions
\begin{equation}
\phi(\vect{X}) = \phi_r (\vect{X}) + \phi_\text{inh}(\vect{X})
\end{equation}
where $\phi_r(\vect{X})=V(R-R_i)/(R_m-R_i)$ if $R\leq R_m$, while $\phi_r(\vect{X})=V$ if $R>R_m$.
Thus, the unknowns of the problem are $\vect{u}$, $\phi_\text{inh}$ and the Lagrange multiplier $\alpha$.
The fields $\vect{u}$ and $\phi_\text{inh}$ are approximated by means of a mixed finite element formulation, using continuous piecewise quadratic functions for the displacement field and continuous piecewise linear functions for $\phi_\text{inh}$. In order to trigger the bifurcation, we apply a small perturbation to the mesh. The expression of the perturbation is given by the critical mode provided by the results of the linear stability analysis \cite{Riccobelli_2017}.

\begin{figure}[b!]
    \centering
    \includegraphics[width=0.7\textwidth]{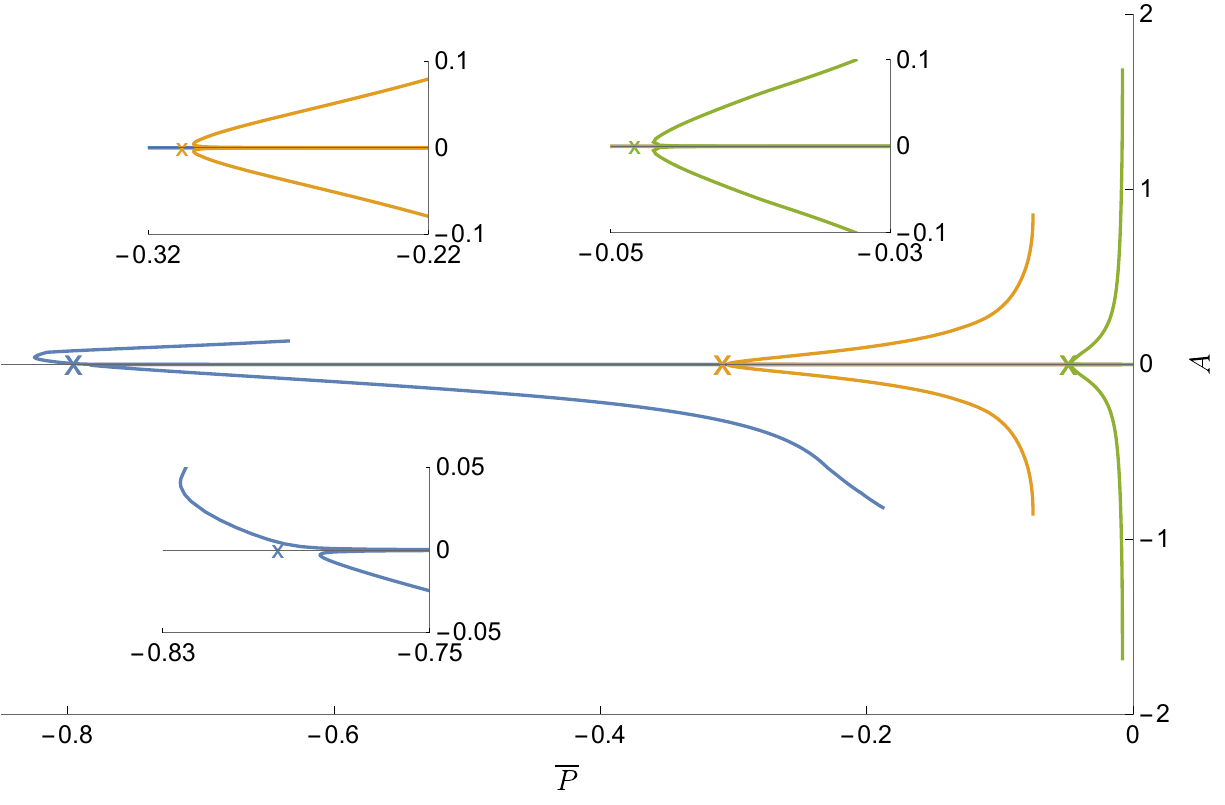}
    \caption{Post-buckling behavior showing $\overline{P}$ versus the dimensionless amplitude  of the deformation at the free surface,  setting $\overline{V}=0.3$, $\mu^e/\mu^d=10$, $t^d=0.9$, and $t^e=0.9,\,0.95, 1$ (blue, orange, and green lines, respectively). The  $\times$ marker denotes the theoretical buckling threshold.}
    \label{fig:bif_diag_comp}
\end{figure}

The numerical scheme is implemented in Python by using the open-source computing platform FEniCS, version 2019.2 \cite{logg2012automated}.  We use PETSc as a linear algebra back-end. In order to reconstruct the bifurcation diagram, we use a pseudo-arclength continuation algorithm \cite{seydel2009practical}, where we use the pressure $P$ as a control parameter of the problem. The nonlinear problem is solved by using a predictor-corrector method. In particular, we adopt a secant predictor to obtain a first guess of the solution and a SNES Newton solver as a corrector. The variational formulation, as well as the Jacobian of the linearized problem, are computed by means of the library UFL \cite{alnaes2014unified}. 
We exploit the library BiFEniCS (\url{https://github.com/riccobelli/bifenics}) for the implementation of the continuation algorithm \cite{riccobelli2021rods}.

\subsection{Results of the numerical simulations}
In the following, we show the results of simulations for the cases analyzed in Fig.~\ref{figure8}. Specifically, we take $t^d=0.9$ and $t^e=0.9,\,0.95,\,1$ (the latter case corresponds to a single DE layer), with $\mu^e/\mu^d=10$ and $\overline{V}=0.3$.

We apply a perturbation of amplitude $\delta A=\pm 10^{-5}R_o$. In Fig.~\ref{fig:bif_diag_comp} we show the bifurcation diagram for $\overline{P}<0$. As a measure of the amplitude of the bifurcated pattern, we use the following scalar dimensionless quantity
\begin{equation}
A = \pm\frac{1}{R_o}\left(\max_{|X|=R_o}|\vect{u}|-\min_{|X|=R_o}|\vect{u}|\right),
\end{equation}
where we take a plus or a minus in front of the amplitude depending on the sign of $\delta A$.

In the cases $t^e=0.95,\,1$, the buckling mode is $m=5$, which is symmetric with respect to the substitution $\delta A\rightarrow -\delta A$. As expected, the bifurcation diagram for this case is symmetric with respect to the $\overline{P}$-axis. In both cases, the shape of the bifurcation is a subcritical pitchfork. Conversely, for $t^e=0.9$ the buckling mode is even. In such a case, the buckling mode is not symmetric with respect to the substitution $\delta A\rightarrow -\delta A$ . As depicted Fig.~\ref{fig:bif_diag_comp}, the bifurcation related to this case becomes transcritical.
In all the cases, the numerical outcomes are in good agreement with the theoretical buckling thresholds, as shown in Fig.~\ref{fig:bif_diag_comp}.

\begin{figure}[t!]
    \centering
    \includegraphics[width=\textwidth]{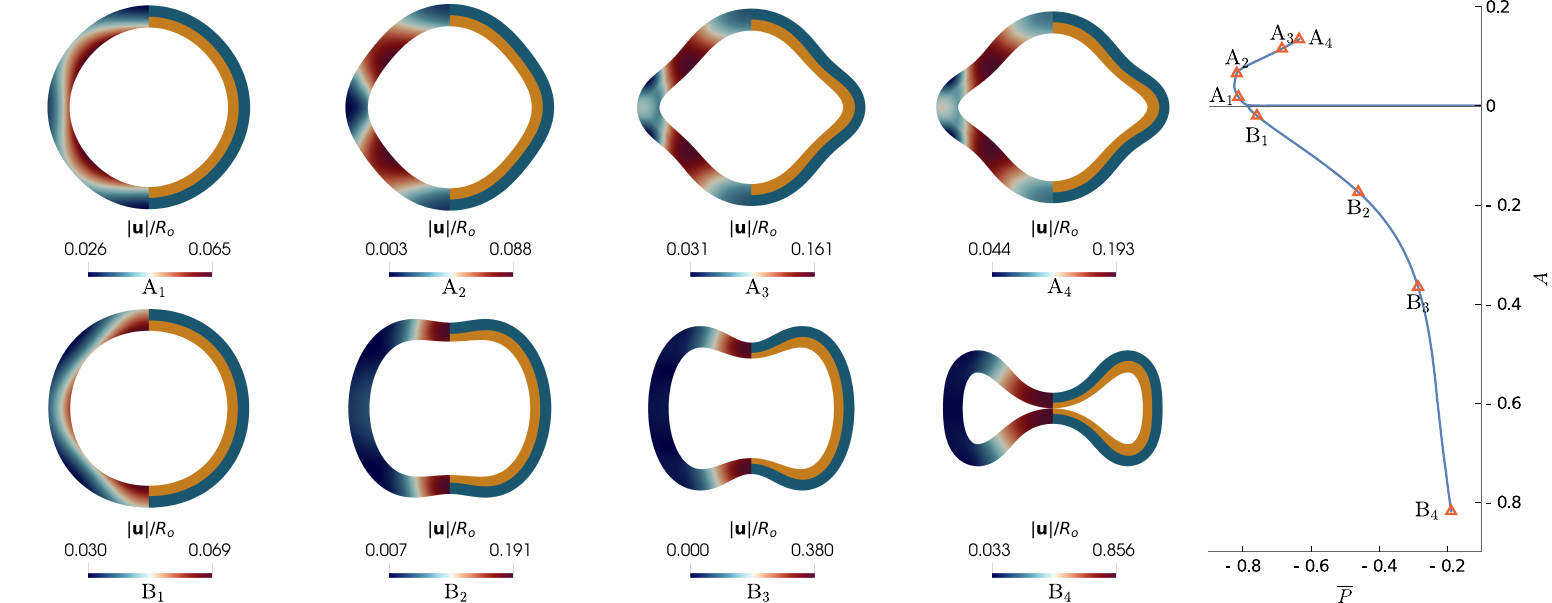}
    \caption{(left) Plot of the actual configuration of the planar section of  spherical bilayers, with $t^d=0.9$, $t^e=0.9$, $\mu^e/\mu^d=10$. Here, we show a section of the shell, where on the left we show $\|\vect{u}\|/R_o$, while on the right each layer is identified using different colors (orange: DE, blue: elastomer). On the right, we show the corresponding points to each configuration on the bifurcation diagram.}
    \label{fig:conf09}
\end{figure}
\begin{figure}[t!]
    \centering
    \includegraphics[width=\textwidth]{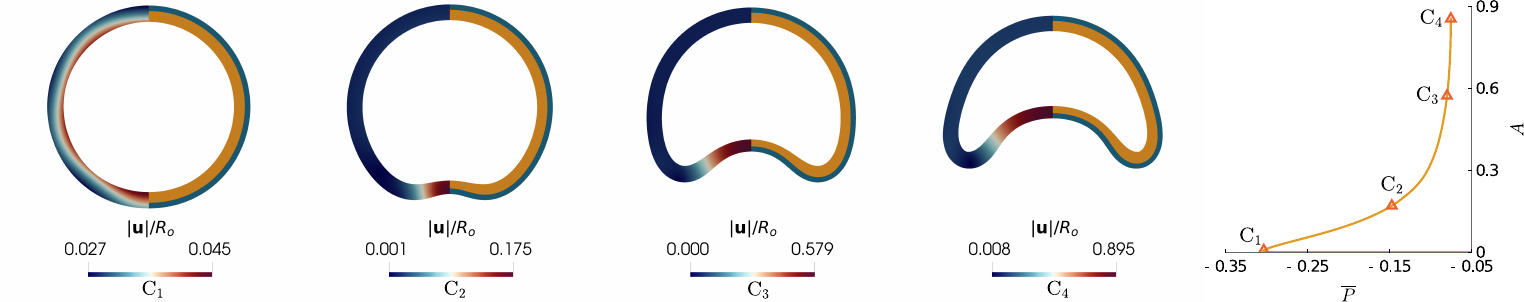}
    \caption{(left) Plot of the actual configuration of the planar section of  spherical bilayers, with $t^d=0.9$, $t^e=0.95$, $\mu^e/\mu^d=10$, $\overline{V}=0.3$. Here, we show a section of the shell, where on the left we show $\|\vect{u}\|/R_o$, while on the right each layer is identified using different colors (orange: DE, blue: elastomer). On the right, we show the corresponding points to each configuration on the bifurcation diagram.}
    \label{fig:conf095}
\end{figure}

In Figs.~\ref{fig:conf09}-\ref{fig:conf095}, we show the evolution of the buckled configurations for the bi-layered balloon. In particular, in the case where $t^e=0.9$ we have two possible buckled configurations: in the former we observe the formation of a protrusion along the equatorial line; while in the latter the two poles collapse until we reach the self-contacting configuration of the shell. Instead, if $t^e=0.95,\,1$ the bifurcation diagram is symmetric, and only one branch will be shown. We observe that the formation of a dimple closes to one of the poles. The initial amplitude of such a dimple is dictated by the critical buckling mode.

\begin{figure}[t!]
    \centering
    \includegraphics[width=0.4\textwidth]{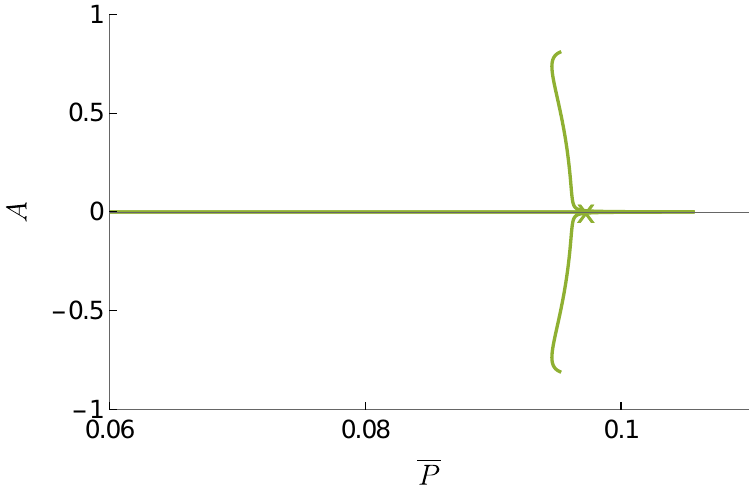}
    \caption{Bifurcation diagram showing $\overline{P}$ versus the dimensionless amplitude of the deformation on the free surface $A$ for $\overline{V}=0.3$, $\mu^e/\mu^d=10$, $t^d=0.9$, and $t^e=1$, $\overline{V}=0.3$. The marker $\times$ denotes the theoretical buckling threshold.}
    \label{fig:bif_diag_ext}
\end{figure}
\begin{figure}[t!]
    \centering
    \includegraphics[width=\textwidth]{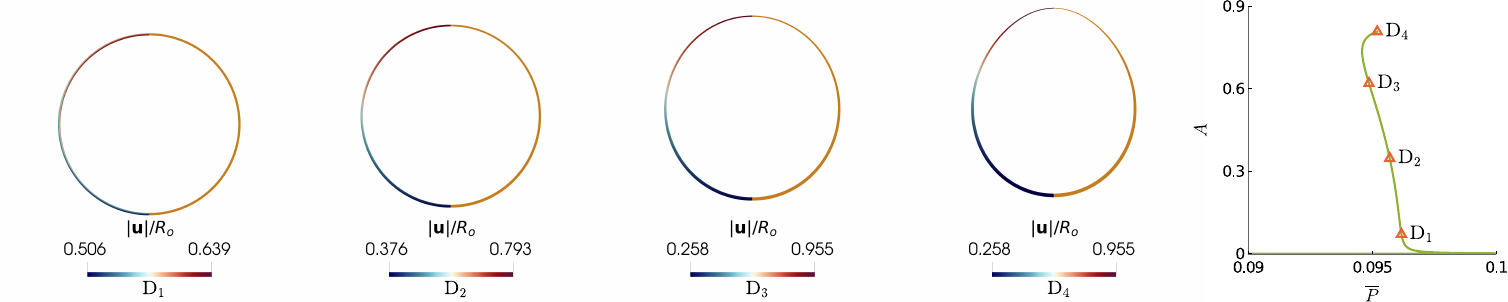}
    \caption{(left) Plot of the actual configuration of  DE  balloons, with $t^d=0.9$, $\mu^e/\mu^d=10$, $\overline{V}=0.3$. Here, we show a planar section of the balloon, where on the left we show $\|\vect{u}\|/R_o$, while on the right each layer is identified using different colors (orange: DE, blue: elastomer). On the right, we show the corresponding points to each configuration on the bifurcation diagram.}
    \label{fig:conf_ext}
\end{figure}

In extension, only the DE balloon ($t^e=1$) undergoes necking. As shown in Fig.~\ref{fig:bif_diag_ext}, the bifurcation is symmetric with respect to the $\overline{P}$-axis, exhibiting  a pitchfork bifurcation at a stretch that is higher than the snap-through threshold, see Fig.~\ref{figure7}. Therefore, $\overline{P}$ first increases beyond the necking threshold, then decreases before undergoing necking. In Fig.~\ref{fig:conf_ext}, we show the actual configuration of the DE balloon. We observe a progressive thinning of the dielectric elastomer, which is a sign of criticality since it is regarded as a precursor of the DE failure \cite{Zurlo_2017}.

\section{Conclusions}
\label{section-conclusion}
DEs are promising electro-mechanical materials, especially suitable for applications as soft actuators and functional wearable devices. The mathematical analysis of layered DE balloons is of considerable complexity, due to many theoretical and numerical challenges given by the geometric and material nonlinearities, as well as the electromechanical coupling. In this paper, we have proposed a theoretical framework for the analysis of the nonlinear response and the bifurcation diagram of layered dielectric-elastic balloons, reporting  complex morphological transitions due to the interplay of  snap-through, buckling, and necking instabilities. After performing a linear stability analysis based on incremental methods in nonlinear elasticity, we have implemented a numerical algorithm using an original mixed finite element  approach coupled with a pseudo-arclength continuation method to investigate the shape transitions of the balloon in the fully nonlinear regime. The onset of the bifurcated branches in the numerical simulations are in excellent agreement with the theoretical marginal stability thresholds. 

In the nonlinear regime,  we found that axisymmetric odd modes result into pitchfork bifurcations for balloons subject to negative inner pressure, while even modes are associated with transcritical bifurcations, as shown in Fig.~\ref{fig:bif_diag_comp}. Not surprisingly, the two branches associated with such transcritical bifurcation exhibit very different behaviors, leading to distinct morphologies, see Fig.~\ref{fig:conf09}.
Conversely, if the shell is inflated we observe necking, where the critical mode is always equal to one. The finite element simulations show a progressive thinning of the DE in the nonlinear regime. This is regarded in the literature as a precursor of failure of the DE \cite{Zurlo_2017}. Such a transition takes place in the unstable region of the snap-through instability, see Fig.~\ref{fig:conf_ext}.

In summary, we have shown that during the snap-through process, a mono-layered DE balloon may be subject to necking, which limits its applications. A layered dielectric-elastic balloon allows to overcome this drawback. Our numerical results also demonstrated that the presence of the elastic layer outside the DE layer has a stabilizing effect on the contractile buckling, and can suppress the necking induced by the snap-through instability of the balloon. 

In this work, we adopted some simplifications that deserve a final discussion. For example, the DE and elastic layers of the balloon are taken to be perfectly bonded. While imperfections may exist in layered structures due to manufacturing problems. However, they have been found of negligible influence on the mechanical response of the structures \cite{Emam2018}. In addition, we did not take into account the influences of viscoelasticity \cite{Plante2006} and electric breakdown failure \cite{Stark1955}, which have been experimentally observed in DEs, and that will be the focus of future studies. As regards the nonlinear finite element analysis, other non-axisymmetric modes could also occur. Future efforts will be devoted to the extension of the proposed numerical scheme to three-dimensional simulations. Furthermore, it would also been interesting to compare our theoretical results with some experiments on spherical DE balloons.

Despite making specific constitutive assumptions for  illustrative purposes, the findings in this paper can be generally applied to give new paradigms for the design and fabrication of functional DE devices. Indeed, we have shown that the presence of an elastic layer can modulate the response of the DE. By tuning the stiffness and the thickness of the elastic layer, we have sown how either  to select the balloon morphological transition or to delay (or even inhibit) the DE necking.

\bigskip

\noindent\textsc{Data Accessibility.} The source code is available on GitHub: \url{https://github.com/riccobelli/dielectric_elastomer_balloon}.

\medskip
\noindent \textsc{Authors’ Contributions.} \textbf{YS}: Conceptualization, Investigation, Validation, Formal analysis, Visualization, Methodology, Writing – original draft, Writing – review $\&$ editing. \textbf{DR}: Investigation, Validation, Formal analysis, Software, Visualization, Methodology, Writing – original draft, Writing – review $\&$ editing. \textbf{YC}: Investigation, Validation, Formal analysis, Visualization, Methodology, Writing – review $\&$ editing. \textbf{WC}: Supervision, Formal analysis, Writing – review $\&$ editing. \textbf{PC}: Conceptualization, Methodology, Writing – original draft, Writing – review $\&$ editing, Funding acquisition, Supervision.

\medskip
\noindent \textsc{Authors’ Contributions.} We declare we have no competing interests.

\medskip
\noindent \textsc{Funding.} YS, DR, and PC have been partially supported by MUR, PRIN Research Projects 2020F3NCPX and grant Dipartimento di Eccellenza 2023-2027,
DR has been partially supported by the National Institute of Higher Mathematics through the grant ``INdAM -- GNFM Project'', code CUP\_E53C22001930001. Partial support from the Shenzhen Scientific and Technological Fund for R\&D, PR China (No. 2021Szvup152) and the National Natural Science Foundation of China (Nos. 12192210 and 12192211) to WC are also acknowledged.

\medskip
\noindent \textsc{Acknowledgements.} YS, DR, and PC are members of the \emph{Gruppo Nazionale di Fisica Matematica} -- INdAM.

\bibliographystyle{abbrv}
\bibliography{refs}

\begin{thebibliography}{10}

\bibitem{Alibakhshi2023}
A.~Alibakhshi, W.~Chen, and M.~Destrade.
\newblock Nonlinear vibration and stability of a dielectric elastomer balloon
  based on a strain-stiffening model.
\newblock {\em Journal of Elasticity}, 153(4-5):533--548, 2023.

\bibitem{alnaes2014unified}
M.~S. Aln{\ae}s, A.~Logg, K.~B. {\O}lgaard, M.~E. Rognes, and G.~N. Wells.
\newblock {Unified Form Language: A domain-specific language for weak
  formulations of partial differential equations}.
\newblock {\em ACM Transactions on Mathematical Software (TOMS)}, 40(2):1--37,
  2014.

\bibitem{atashipour2016electro}
S.~Atashipour and R.~Sburlati.
\newblock Electro-elastic analysis of a coated spherical piezoceramic sensor.
\newblock {\em Composite Structures}, 156:399--409, 2016.

\bibitem{Bertoldi2011}
K.~Bertoldi and M.~Gei.
\newblock Instabilities in multilayered soft dielectrics.
\newblock {\em Journal of the Mechanics and Physics of Solids}, 59(1):18--42,
  2011.

\bibitem{Bortot17}
E.~Bortot.
\newblock Analysis of multilayer electro-active spherical balloons.
\newblock {\em Journal of the Mechanics and Physics of Solids}, 101:250--267,
  2017.

\bibitem{Bortot18}
E.~Bortot and G.~Shmuel.
\newblock Prismatic bifurcations of soft dielectric tubes.
\newblock {\em International Journal of Engineering Science}, 124:104--114,
  2018.

\bibitem{Bustamante_2008}
R.~Bustamante, A.~Dorfmann, and R.~W. Ogden.
\newblock Nonlinear electroelastostatics: a variational framework.
\newblock {\em Zeitschrift für angewandte Mathematik und Physik},
  60(1):154--177, mar 2008.

\bibitem{Chen2017}
G.~Chen.
\newblock {\em Controlling Chaos and Bifurcations in Engineering Systems}.
\newblock CRC press, 1999.

\bibitem{debotton2013axisymmetric}
G.~DeBotton, R.~Bustamante, and A.~Dorfmann.
\newblock Axisymmetric bifurcations of thick spherical shells under inflation
  and compression.
\newblock {\em International Journal of Solids and Structures}, 50(2):403--413,
  2013.

\bibitem{Dorfmann2010}
A.~Dorfmann and R.~W. Ogden.
\newblock Electroelastic waves in a finitely deformed electroactive material.
\newblock {\em IMA Journal of Applied Mathematics}, 75(4):603--636, 2010.

\bibitem{Dorfmann2014}
L.~Dorfmann and R.~W. Ogden.
\newblock Nonlinear response of an electroelastic spherical shell.
\newblock {\em International Journal of Engineering Science}, 85:163--174,
  2014.

\bibitem{dorfmann2014nonlinear}
L.~Dorfmann and R.~W. Ogden.
\newblock {\em Nonlinear Theory of Electroelastic and Magnetoelastic
  Interactions}, volume~1.
\newblock Springer, 2014.

\bibitem{Emam2018}
S.~A. Emam, M.~A. Eltaher, M.~E. Khater, and W.~S. Abdalla.
\newblock Postbuckling and free vibration of multilayer imperfect nanobeams
  under a pre-stress load.
\newblock {\em Applied Sciences}, 8(11):2238, 2018.

\bibitem{Gent99}
A.~N. Gent and I.~S. Cho.
\newblock Surface instabilities in compressed or bent rubber blocks.
\newblock {\em Rubber Chemistry and Technology}, 72(2):253--262, 1999.

\bibitem{Godaba2014}
H.~Godaba, C.~C. Foo, Z.~Q. Zhang, B.~C. Khoo, and J.~Zhu.
\newblock Giant voltage-induced deformation of a dielectric elastomer under a
  constant pressure.
\newblock {\em Applied Physics Letters}, 105(11):112901, 2014.

\bibitem{haughton1980post}
D.~M. Haughton.
\newblock Post-bifurcation of perfect and imperfect spherical elastic
  membranes.
\newblock {\em International Journal of Solids and Structures},
  16(12):1123--1133, 1980.

\bibitem{Haughton1978}
D.~M. Haughton and R.~W. Ogden.
\newblock On the incremental equations in non-linear elasticity—ii.
  bifurcation of pressurized spherical shells.
\newblock {\em Journal of the Mechanics and Physics of Solids}, 26(2):111--138,
  1978.

\bibitem{hill1975bifurcation}
R.~Hill and J.~W. Hutchinson.
\newblock Bifurcation phenomena in the plane tension test.
\newblock {\em Journal of the Mechanics and Physics of Solids},
  23(4-5):239--264, 1975.

\bibitem{Huang2013}
J.~Huang, S.~Shian, Z.~Suo, and D.~R. Clarke.
\newblock Maximizing the energy density of dielectric elastomer generators
  using equi-biaxial loading.
\newblock {\em Advanced Functional Materials}, 23(40):5056--5061, 2013.

\bibitem{hutchinson1974bifurcation}
J.~W. Hutchinson and J.~P. Miles.
\newblock Bifurcation analysis of the onset of necking in an elastic/plastic
  cylinder under uniaxial tension.
\newblock {\em Journal of the Mechanics and Physics of Solids}, 22(1):61--71,
  1974.

\bibitem{Jin2017}
X.~Jin and Z.~Huang.
\newblock Random response of dielectric elastomer balloon to electrical or
  mechanical perturbation.
\newblock {\em Journal of Intelligent Material Systems and Structures},
  28(2):195--203, 2017.

\bibitem{Kim2012}
B.~Kim, S.~B. Lee, J.~Lee, S.~Cho, H.~Park, S.~Yeom, and S.~H. Park.
\newblock A comparison among neo-hookean model, mooney-rivlin model, and ogden
  model for chloroprene rubber.
\newblock {\em International Journal of Precision Engineering and
  Manufacturing}, 13(5):759--764, 2012.

\bibitem{Kim07}
K.~J. Kim and S.~Tadokoro.
\newblock Electroactive polymers for robotic applications.
\newblock {\em Artificial Muscles and Sensors}, 23:291, 2007.

\bibitem{Kumar2023}
A.~Kumar, A.~Khurana, A.~K. Patra, Y.~Agrawal, and M.~M. Joglekar.
\newblock Electromechanical performance of dielectric elastomer composites:
  Modeling and experimental characterization.
\newblock {\em Composite Structures}, 320:117130, 2023.

\bibitem{Lee2022}
D.-Y. Lee, S.~H. Jeong, A.~J. Cohen, D.~M. Vogt, M.~Kollosche, G.~Lansberry,
  Y.~Meng{\"u}{\c{c}}, A.~Israr, D.~R. Clarke, and R.~J. Wood.
\newblock A wearable textile-embedded dielectric elastomer actuator haptic
  display.
\newblock {\em Soft Robotics}, 2022.

\bibitem{Li2013}
T.~Li, C.~Keplinger, R.~Baumgartner, S.~Bauer, W.~Yang, and Z.~Suo.
\newblock Giant voltage-induced deformation in dielectric elastomers near the
  verge of snap-through instability.
\newblock {\em Journal of the Mechanics and Physics of Solids}, 61(2):611--628,
  2013.

\bibitem{Liang2015}
X.~Liang and S.~Cai.
\newblock Shape bifurcation of a spherical dielectric elastomer balloon under
  the actions of internal pressure and electric voltage.
\newblock {\em Journal of Applied Mechanics}, 82(10):101002, 2015.

\bibitem{logg2012automated}
A.~Logg, K.-A. Mardal, and G.~Wells.
\newblock {\em Automated Solution of Differential Equations by the Finite
  Element Method: The FEniCS Book}, volume~84.
\newblock Springer Science \& Business Media, 2012.

\bibitem{Mao2019}
R.~Mao, B.~Wu, E.~Carrera, and W.~Chen.
\newblock Electrostatically tunable small-amplitude free vibrations of
  pressurized electro-active spherical balloons.
\newblock {\em International Journal of Non-Linear Mechanics}, 117:103237,
  2019.

\bibitem{martin2020stroh}
P.~Martin.
\newblock A stroh formalism for small-on-large problems in spherical polar
  coordinates.
\newblock {\em Journal of Elasticity}, 138(2):125--144, 2020.

\bibitem{Melnikov2020}
A.~Melnikov, L.~Dorfmann, and R.~W. Ogden.
\newblock Bifurcation of finitely deformed thick-walled electroelastic
  spherical shells subject to a radial electric field.
\newblock {\em International Journal of Non-Linear Mechanics}, 121:103429,
  2020.

\bibitem{Norris_2010}
A.~N. Norris and A.~L. Shuvalov.
\newblock Wave impedance matrices for cylindrically anisotropic radially
  inhomogeneous elastic solids.
\newblock {\em The Quarterly Journal of Mechanics and Applied Mathematics},
  63(4):401--435, jul 2010.

\bibitem{Halloran2008}
A.~O’Halloran, F.~O’malley, and P.~McHugh.
\newblock A review on dielectric elastomer actuators, technology, applications,
  and challenges.
\newblock {\em Journal of Applied Physics}, 104(7):9, 2008.

\bibitem{Pang2020}
W.~Pang, X.~Cheng, H.~Zhao, X.~Guo, Z.~Ji, G.~Li, Y.~Liang, Z.~Xue, H.~Song,
  F.~Zhang, et~al.
\newblock Electro-mechanically controlled assembly of reconfigurable 3d
  mesostructures and electronic devices based on dielectric elastomer
  platforms.
\newblock {\em National science review}, 7(2):342--354, 2020.

\bibitem{Plante2006}
J.-S. Plante and S.~Dubowsky.
\newblock Large-scale failure modes of dielectric elastomer actuators.
\newblock {\em International journal of solids and structures},
  43(25-26):7727--7751, 2006.

\bibitem{Riccobelli_2017}
D.~Riccobelli and P.~Ciarletta.
\newblock Shape transitions in a soft incompressible sphere with residual
  stresses.
\newblock {\em Mathematics and Mechanics of Solids}, 23(12):1507--1524, dec
  2017.

\bibitem{riccobelli2021rods}
D.~Riccobelli, G.~Noselli, and A.~DeSimone.
\newblock Rods coiling about a rigid constraint: helices and perversions.
\newblock {\em Proceedings of the Royal Society A}, 477(2246):20200817, 2021.

\bibitem{Rudykh2009}
S.~Rudykh, K.~Bhattacharya, and G.~Debotton.
\newblock Snap-through actuation of thick-wall electroactive balloons.
\newblock {\em International Journal of Non-Linear Mechanics}, 47(2):206--209,
  2012.

\bibitem{seydel2009practical}
R.~Seydel.
\newblock {\em Practical Bifurcation and Stability Analysis}, volume~5.
\newblock Springer Science \& Business Media, 2009.

\bibitem{Sharma2018}
A.~K. Sharma, N.~Arora, and M.~M. Joglekar.
\newblock Dc dynamic pull-in instability of a dielectric elastomer balloon: an
  energy-based approach.
\newblock {\em Proceedings of the Royal Society A: Mathematical, Physical and
  Engineering Sciences}, 474(2211):20170900, 2018.

\bibitem{Stark1955}
K.~H. Stark and C.~G. Garton.
\newblock Electric strength of irradiated polythene.
\newblock {\em Nature}, 176(4495):1225--1226, 1955.

\bibitem{Su2019a}
Y.~Su, W.~Chen, and M.~Destrade.
\newblock Tuning the pull-in instability of soft dielectric elastomers through
  loading protocols.
\newblock {\em International Journal of Non-Linear Mechanics}, 113:62--66,
  2019.

\bibitem{Su2018}
Y.~Su, B.~Wu, W.~Chen, and C.~L{\"u}.
\newblock Optimizing parameters to achieve giant deformation of an
  incompressible dielectric elastomeric plate.
\newblock {\em Extreme Mechanics Letters}, 22:60--68, 2018.

\bibitem{Su2020}
Y.~Su, B.~Wu, W.~Chen, and D.~Michel.
\newblock Pattern evolution in bending dielectric-elastomeric bilayers.
\newblock {\em Journal of the Mechanics and Physics of Solids}, 136:103670,
  2020.

\bibitem{Suo08}
Z.~Suo, X.~Zhao, and W.~H. Greene.
\newblock A nonlinear field theory of deformable dielectrics.
\newblock {\em Journal of the Mechanics and Physics of Solids}, 56(2):467--486,
  2008.

\bibitem{Xie2016}
Y.~Xie, J.~Liu, and Y.~Fu.
\newblock Bifurcation of a dielectric elastomer balloon under pressurized
  inflation and electric actuation.
\newblock {\em International Journal of Solids and Structures}, 78:182--188,
  2016.

\bibitem{Xu2020}
F.~Xu, C.~Fu, and Y.~Yang.
\newblock Water affects morphogenesis of growing aquatic plant leaves.
\newblock {\em Physical Review Letters}, 124(3):038003, 2020.

\bibitem{Zhao2007}
X.~Zhao and Z.~Suo.
\newblock Method to analyze electromechanical stability of dielectric
  elastomers.
\newblock {\em Applied Physics Letters}, 91(6):061921, 2007.

\bibitem{Zhu10}
J.~Zhu, H.~Stoyanov, G.~Kofod, and Z.~Suo.
\newblock Large deformation and electromechanical instability of a dielectric
  elastomer tube actuator.
\newblock {\em Journal of Applied Physics}, 108(7):074113, 2010.

\bibitem{Zurlo_2017}
G.~Zurlo, M.~Destrade, D.~DeTommasi, and G.~Puglisi.
\newblock Catastrophic thinning of dielectric elastomers.
\newblock {\em Physical Review Letters}, 118(7), feb 2017.

\end{thebibliography}
\end{document}